\documentclass[twocolumn,showpacs,prb,amsmath,amssymb]{revtex4}

\usepackage{graphicx}
\usepackage{bm}

\begin{document}

\title{Temperature-dependent magnetization in
  diluted magnetic semiconductors}

\author{S. \surname{Das Sarma}}
\author{E.H. Hwang}
\author{A. Kaminski}
\affiliation{Condensed Matter Theory Center,
Department of Physics, University of Maryland, College Park, Maryland
20742-4111}

\begin{abstract}
We calculate magnetization in magnetically doped semiconductors
assuming a local exchange model of carrier-mediated ferromagnetic
mechanism and using a number of complementary theoretical approaches.
In general, we find that the results of our mean-field calculations,
particularly the dynamical mean field theory results, give excellent
qualitative agreement with the experimentally observed magnetization
in systems with itinerant charge carriers, such as Ga$_{1-x}$Mn$_x$As
with $0.03 < x < 0.07$, whereas our percolation-theory-based
calculations agree well with the existing data in strongly insulating
materials, such as Ge$_{1-x}$Mn$_x$.  We comment on the issue of
non-mean-field like magnetization curves and on the observed
incomplete saturation magnetization values in diluted magnetic
semiconductors from our theoretical perspective.  In agreement with
experimental observations, we find the carrier density to be the
crucial parameter determining the magnetization behavior.  Our
calculated dependence of magnetization on external magnetic field is
also in excellent agreement with the existing experimental data.
\end{abstract}


\maketitle

\section{Introduction}

Diluted magnetic semiconductors (sometimes also referred to as doped
magnetic semiconductors -- we will use the abbreviation ``DMS'' to
denote both of these equivalent terminologies) have recently attracted
a great deal of attention\cite{sds1,new2} for their potential in
combining ferromagnetic and semiconductor properties in a single
material. The prototypical DMS material is
Ga$_{1-x}$Mn$_x$As\cite{sds13} (typically $x\approx 1-10$\%) with the
Mn ions substitutionally (in the ideal situation) replacing Ga at the
cation sites. Mn ions in Ga$_{1-x}$Mn$_x$As serve a dual purpose,
acting both as dopants (acceptors in this case) and as magnetic
impurities, whose spins align at the ferromagnetic transition. For
$x\approx 1-7$\%, Ga$_{1-x}$Mn$_x$As is found to be ferromagnetic with
the ferromagnetic transition temperature (or, equivalently, the Curie
temperature) $T_c\approx 10 - 100$K. The optimum value of $x$, which
corresponds to the highest value of reported value of $T_c$, is around
5\%. Other DMS materials of current interest include
In$_{1-x}$Mn$_x$As,\cite{new4} Ga$_{1-x}$Mn$_x$P,\cite{new5}
Ge$_{1-x}$Mn$_x$,\cite{new6} and Ga$_{1-x}$Mn$_x$Sb.\cite{new7}

In spite of a great deal of recent experimental and theoretical
activity,\cite{sds1,new2} there is not yet a consensus on the
fundamental mechanism leading to ferromagnetism in these systems as
well as on the definitive predictive theory quantitatively describing
this ferromagnetic mechanism. It is therefore important to work out
detailed and experimentally falsifiable consequences for various
proposed theoretical models and ideas. In this context, it is
unfortunate that much of the theoretical DMS literature, perhaps due
to the considerable technological motivation in creating
room-temperature ferromagnetic semiconductors for projected
spintronics applications, has concentrated on the calculations of
$T_c$ in various DMS materials. Such theoretical predictions of $T_c$
invariably involve tuning free parameters ({\it e.~g.}, the strength
of the exchange coupling between carriers and local moments), whose
values are often unknown \emph{a priori}. This significantly reduces
the practical importance of these predictions except perhaps in the
broadest qualitative sense of identifying the crucial controlling
parameters which determine and limit $T_c$ in DMS materials. On the
other hand, the temperature dependence $M(T)$ of the spontaneous
magnetization possesses many characteristics, such as
concavity/convexity of the curve, value of saturation magnetization,
critical behavior at the point of ferromagnetic transition \emph
{etc.}, which cannot all be described by just tuning parameters of a
given model. Thus study of $M(T)$ has a very high potential for
elucidating the physics behind DMS ferromagnetism in real systems.

A particular issue of considerable significance in DMS ferromagnetism
has been the non-trivial non-mean-field-like behavior of the
spontaneous magnetization as a function of temperature. This was
already apparent in the very first reported observation\cite{new4} of
DMS ferromagnetism in a III$_{1-x}$Mn$_x$V material, namely
In$_{1-x}$Mn$_x$As, where the experimental $M(T)$ curve exhibited an
untypical outwardly concave shape strikingly different from the usual
convex $M(T)$ behavior expected within the textbook Weiss mean-field
theory\cite{sds14} and seen routinely in conventional ferromagnetic
materials. The In$_{1-x}$Mn$_x$As (with $x=0.013$) system studied in
Ref.~\onlinecite{new4} was an insulating system (\emph{i.\ e.}\ with
resistivity increasing monotonically with decreasing temperature), and
the insulating ferromagnetic DMS systems studied so far in the
literature almost always exhibit qualitatively similar
non-mean-field-like concave $M(T)$ behavior.\cite{new6}

Two of us have recently shown \cite{sds7} that such a manifestly
non-mean-field-like concave $M(T)$ behavior in insulating DMS
materials can be understood on the basis of a magnetic percolation
transition of bound magnetic polarons in the strongly localized
carrier system. Earlier numerical simulations \cite{sds15,sds16} in
the strongly localized regime had already indicated that the $M(T)$
behavior of DMS ferromagnets could have concave outward shapes as seen
experimentally.  Very recent numerical simulations \cite{sds17} in the
strongly localized regime have verified our polaron percolation picture of
DMS ferromagnetism in the insulating regime.

Even in the ``metallic'' (\emph{i.\ e.}\ with resistivity decreasing
with $T$ below the ferromagnetic transition temperature) DMS systems,
\cite{sds1,sds13,new13,new14,new15,new16,new17,new18,new19,sds19} such
as optimally doped Ga$_{0.95}$Mn$_{0.05}$As, where long range magnetic
ordering of the Mn magnetic moments is created presumably by a dilute
gas of delocalized holes mediating the magnetic interaction, the
experimentally observed $M(T)$ often appears to be very different from
the classic mean-field shape.\cite{sds14} Although the ``metallic''
Ga$_{1-x}$Mn$_x$As DMS system does not exhibit the manifestly concave
temperature dependent magnetization seen in insulating DMS systems,
the observed $M(T)$ in metallic DMS is often
\cite{sds1,sds13,new13,new14,new15,new16,new17,new18,new19,sds19}
almost linear in temperature for $0.5 T_c \lesssim T < T_c$, with a
temperature-dependent behavior intermediate between the concave $M(T)$
shapes of the localized theory\cite{sds7} and the textbook convex
magnetization curves.\cite{sds14}

We mention that very recent annealing experiments
\cite{new13,new14,new15,new16,new17,new18,new19,sds19} in
Ga$_{1-x}$Mn$_x$As (with $x=0.05 - 0.10$) demonstrate that $M(T)$
behavior (as well as the value of $T_c$) can be strongly affected by
annealing, and in particular, ``optimal'' annealing (to be found
empirically for each sample purely by trial and error since the
precise role of annealing in improving the materials quality is
unknown at this stage) may lead to reasonably mean-field-like convex
$M(T)$ shape with a concomitant enhancement of $T_c$.  ``Non-optimal''
annealing, on the other hand, leads to suppression of $T_c$ and
strongly non-mean-field-like (and non-universal) $M(T)$ behavior.

Finally, most of DMS magnetization measurements in the literature,
particularly for the insulating DMS systems, show a saturation
magnetization (for $B=0$) considerably smaller in magnitude than that
expected from the full ordering of all the magnetic ions, indicating
that a large fraction (sometimes as much as 90\%) of the magnetic
ions do not contribute to the global DMS ferromagnetism.

Motivated by the desire for illuminating the physical mechanisms
underlying DMS ferromagnetism, we theoretically consider in this paper
the temperature-dependent magnetization $M(T)$ in three-dimensional
DMS systems using a number of complementary theoretical approaches.
The calculations we present here are based on the static (Weiss)
molecular mean-field theory,\cite{new21} the dynamical mean-field
theory (DMFT),\cite{sds6} and the percolation
theory.\cite{EfrosShklovskiiBook} Of particular interest is the
important issue of correlations between magnetic and transport
properties of various DMS materials since such correlations should
(and do) exist in magnetic systems where the ferromagnetism is
mediated by carriers leading to the global ordering of the dopant
local moments.  While earlier studies invariably concentrated on
either localized-carrier (``insulating'') and itinerant-carrier
(``metallic'') material, we consider both these regimes as well as the
crossover between them.  We provide detailed numerical results for our
calculated magnetization $M(T,B)$, where $B$ is the applied external
magnetic field and $T$ the temperature, in various regimes of the
system parameter space.  These $M(T,B)$ results should help ascertain
the applicability of various theoretical models to specific
experimental DMS materials of current interest.

This paper is organized as follows. In Sec.~\ref{sec:2}, we describe
our model and our theoretical approaches based on the Weiss mean-field
theory, DMFT, and the percolation theory, and present our numerical
results for $M(T,B)$ in each of the theoretical approaches providing
brief discussion of our results in light of the existing experimental
data in the literature.  We conclude in Sec.~\ref{sec:3} summarizing
our qualitative findings and providing a critical perspective on what
our theoretical results on magnetization imply for the microscopic
mechanisms underlying DMS ferromagnetism with the particular emphasis
on the correlations between transport characteristics (``metallic'' or
``insulating'') and $M(T)$ behavior (convex, concave, or ``linear'').


\section{Model, Theory, and Results}\label{sec:2}

\subsection{The model}

We assume in this work that the fundamental mechanism underlying
ferromagnetism in DMS materials (\emph{e.~g.} III$_{1-x}$Mn$_x$V,
Ge$_{1-x}$Mn$_x$) is the carrier -- local-moment (kinetic or $p$-$d$)
exchange coupling, which eventually leads to a global ferromagnetic
ordering of the impurity local moments (\emph{i.~e.} Mn) for $T<T_c$
overcoming any direct antiferromagnetic (super-exchange) interaction
between the local moment spins themselves. This is certainly the
prevalent viewpoint for DMS ferromagnetism, at least for
Ga$_{1-x}$Mn$_x$As system following the pioneering work of Ohno and
his collaborators. It is natural to ask about the evidence for this
belief in carrier-mediated DMS ferromagnetism induced in the Mn local
moments. First-principles band theory calculations indicate that there
is strong $p$-$d$ hybridization between the Mn $d$-levels and the
valence band $p$-states of GaAs. This leads to strong kinetic exchange
coupling between hole spins and Mn spins, which is the basis of our
model. Experimentally, there are a number of compelling circumstantial
indications of Ga$_{1-x}$Mn$_x$As being a carrier-mediated
ferromagnetic material. First, there is the strong correlation (made
even stronger by the recent annealing experiments) between transport
and magnetic properties. This, however, is not definitively conclusive
since strongly insulating Ga$_{1-x}$Mn$_x$As (and In$_{1-x}$Mn$_x$As)
samples are also found to be ferromagnetic (albeit with much lower
$T_c$ values). Perhaps the most compelling evidence supporting the
carrier-mediated ferromagnetic mechanism is the observed agreement (so
far available only in Ga$_{1-x}$Mn$_x$As DMS systems) between the
magnetization measured directly using a SQUID magnetometer and that
inferred by analyzing the anomalous Hall effect data. Such an
consistency between direct and ``transport-inferred'' magnetization
strongly suggests a carrier-mediated exchange mechanism underlying DMS
ferromagnetism. In addition, there are a number of experiments dealing
with optical properties (again, available only for Ga$_{1-x}$Mn$_x$As
so far) which indicate very strong correlation between the magnetic
band structure and the magnetization of the system, leading again to
the condition that the global ferromagnetic ordering of the Mn local
moments is most likely induced by the hole spin polarization in DMS
systems. The related issue of whether these carriers are valence band
holes or impurity band holes is more difficult to settle. Given
the strong inherent disorder in DMS materials and rather strong
exchange coupling between hole and Mn spins (leading to local
binding of holes and Mn ions), it is natural to think that all of the
physics (both localized and delocalized) are essentially impurity band
physics. This viewpoint, which we have adopted in our theories
presented in this paper, has received strong support from recent
optical experiments. Also our DMFT results, as presented in
Sec.~\ref{sec:2b}, clearly demonstrate the important role of impurity
band physics in DMS magnetic properties.

Since the most intensively studied DMS materials of current interest
are Mn-doped III-V semiconductors III$_{1-x}$Mn$_x$V (\emph{e.\ g.}\ 
Ga$_{1-x}$Mn$_x$As, In$_{1-x}$Mn$_x$As, Ga$_{1-x}$Mn$_x$P,
Ga$_{1-x}$Mn$_x$N, Ga$_{1-x}$Mn$_x$Sb), where the carriers are
typically holes, we will other refer to the semiconductor carriers as
``holes'' in the rest of the article without any loss of generality.

So our basic model is that of a finite density $n_{\textrm{i}}$ of
magnetic dopants (``impurities'') interacting through a local exchange
coupling with a finite density $n_{\textrm{c}}$ of holes in the host
semiconductor material with $n_{\textrm{c}}/ n_{\textrm{i}}\ll 1$. We
assume that magnetic impurities under consideration enter
substitutionally at the cation sites (\emph{e.~g.} Mn impurities at Ga
sites).  Recent experimental annealing studies of Ga$_{1-x}$Mn$_x$As
have shown \cite{new13,new14,new15,new16,new17,new18,new19} that
lattice defects may be playing an important role in determining
magnetic and transport properties of the samples, but we assume in our
work that these defects enter our theory only in determining the basic
parameters of the model, namely, the density of magnetically active
dopants $n_{\textrm{i}}$, the hole density $n_{\textrm{c}}$, and
perhaps the local effective exchange coupling $J$ between the holes
and the magnetic impurities, and do not include any defects into our
model explicitly.  Examples of such defects which enter our model
through its parameters rather than directly are given by antisite
defects in the host semiconductor material (\emph{i.~e.} As at Ga
sites in GaAs) and Mn interstitials (\emph{i.\ e.}\ Mn atoms at
interstitials sites rather than cation substitutional sites), which
may very well be important in providing substantial compensation in
the semiconductor, leading to the experimental fact that the hole
density $n_{\textrm{c}}$ is usually a small fraction of the magnetic
dopant density instead of there being a one-to-one correspondence
between dopants and holes. The local moment density $n_{\textrm{i}}$
in our model is not necessarily the total Mn concentration in the
system, since the presence of Mn interstitial defects could lower the
density of magnetically active Mn ions. In this work we use
$n_{\textrm{i}}$ to denote local moment volume density and $x$ to
denote the fraction of Ga atoms replaced by Mn dopants.

Similarly, consistent with the spirit of our minimal model we also
neglect all band structure effects in our theory, making the simplest
approximation such as a single parabolic band with a single effective
mass or a single simple tight-binding carrier band characterized by an
effective band width parameter. This is not because realistic band
structure effects are not of any importance in DMS ferromagnetism, in
fact we believe that spin-orbit coupling in Ga$_{1-x}$Mn$_x$As valence
band hole states may play a quantitative role in Ga$_{1-x}$Mn$_x$As
ferromagnetism, particularly in the relatively disorder-free
``metallic'' ($x\approx 0.05$) systems where the holes are
likely to be GaAs valence band hole states with strong spin-orbit
coupling. Our reasons for neglecting spin-orbit coupling and other
band structure effects in our theory are the following: (1) our
interest in this paper is in theoretically exploring $M(T,B)$ within a
minimal model, which requires only $n_{\textrm{i}}$, $n_{\textrm{c}}$,
and $J$, leaving out all non-essential complications; (2) if needed,
band structure effects can be systematically included in the future by
appropriately extending the model; (3) due to the inevitable presence
of strong exchange coupling and strong disorder in DMS materials, a
starting point based on perfect ``realistic'' valence band hole states
may be inapplicable -- in fact, we believe that much of the DMS
physics is occurring in the impurity band of the host semiconductor
with the itinerant and the localized carriers being the extended and
the localized hole states in the impurity band of the system and not
the valence band states, a view point strongly supported by several
recent experimental results.\cite{new23,new24,new25,new26}

The Hamiltonian of exchange interaction between magnetic impurities
and holes we use in this paper reads
\begin{equation}
H_M=\int d^3r \sum_{j}Ja_0^3
\left({\bf S}_j \cdot {\bf s}(\textbf{r})\right)
\delta(\textbf{r}-\textbf{R}_j)
\;,
\label{ham}
\end{equation} 
where $J$, which has units of energy, is the exchange coupling between
an impurity spins ${\bf S}_j$ located at ${\bf R}_j$ and a hole spin
density ${\bf s}({\bf r})$, and $a_0^3$ is the unit cell volume needed
for proper normalization.  The impurity spin {\bf S} in
Eq.~(\ref{ham}) is assumed to be completely classical in the theory
whereas the carrier spin {\bf s} is treated quantum-mechanically.
This is justified because the impurity spin is large, \emph{e.\ g.}\ 
$S=5/2$ for Mn in Ga$_{1-x}$Mn$_x$As.

Our model (except for the mean-field theory considerations of
Sec.~\ref{sec:2a}) omits the direct Mn$-$Mn antiferromagnetic exchange
interaction assuming its effects to be either negligibly small or
incorporated into the effective parameters of the model.  Actually, in
the parameter range of interest to us ($x \ll 1$), where DMS
ferromagnetism typically occurs, the magnetic impurities are separated
from each other by non-magnetic atoms, and this antiferromagnetic
interaction, which rapidly decays with the distance, should be
negligible.  We also ignore any specific hole-hole interaction effect
in our theory.  These approximations are nonessential and are done in
the spirit of identifying the minimal DMS magnetic model of interest.
Both of these effects, which may be of quantitative importance in some
situations, can be included in the theory by adjusting the parameters
of the model or perhaps at the cost of introducing more unknown
parameters characterizing these interactions.  The possible effects of
including these interactions in our calculations will be discussed
later in the paper.

With this introduction the full Hamiltonian is given by
\begin{eqnarray}
H & = &
\int d^3r \sum_{\alpha} \psi^{\dagger}_{\alpha}({\bf r}) \left [
-\frac{\nabla^2} 
{2m} + V({\bf r}) \right ]
\psi^{\vphantom{\dagger}}_{\alpha}({\bf r}) \nonumber \\ 
  & + & \sum_j \int d^3 r 
\left [ \sum_{\alpha} W({\bf r}-{\bf R}_j)  
\psi_{\alpha}^{\dagger}({\bf r}) \psi_{\alpha}({\bf r}) \right.
\nonumber\\
&&\quad+\left.\sum_{\alpha \beta}
Ja_0^3\left({\bf S}_j\cdot \bm{\sigma}_{\alpha \beta}\right) 
\delta({\bf r}-\textbf{R}_j)
\psi_{\alpha}^{\dagger}({\bf r}) 
\psi^{\vphantom{\dagger}}_{\beta}({\bf r})  \right ]\nonumber\\
&+& \sum_j g_{\textrm{i}}\mu_B \left({\bf S}_j\cdot\mathbf{B}\right)\nonumber\\
&+& 
\int d^3 r \sum_{\alpha \beta}g_{\textrm{c}}\mu_B
\psi_{\alpha}^{\dagger}({\bf r}) 
\psi^{\vphantom{\dagger}}_{\beta}({\bf r})
\left(\bm{\sigma}_{\alpha \beta}\cdot\mathbf{B}\right) 
\:,
\label{ham_full}
\end{eqnarray}
where the first term is the ``band'' Hamiltonian with $m$ being the
relevant effective mass, $V({\bf r})$ is the random potential arising
from (non-magnetic) disorder, $W({\bf r}-\bm{R}_i)$ is the Coulomb
potential due to a magnetic impurity located at $\bm{R}_i$, and the
second term is the local exchange coupling between the moments of
magnetic impurities and the carrier spins [\emph{i.\ e.}\ precisely
the $H_M$ defined through Eq.~(\ref{ham})], with
$\bm{\sigma}_{\alpha\beta}$ being the Pauli matrices. The last two
terms in Eq.~(\ref{ham_full}) are simply the Zeeman energies of the
local moments and the holes respectively, $g_{\textrm{i}}$ and
$g_{\textrm{c}}$ are the corresponding $g$-factors, $\mu_B$ is the
Bohr magneton, and $\mathbf{B}$ is the external magnetic field.

The local exchange coupling can be ferromagnetic ($J<0$) or
antiferromagnetic ($J>0$) without affecting DMS ferromagnetism [first
principles band theory \cite{new27,RE,new29} suggests local
antiferromagnetic coupling ($J>0$) for the holes in
Ga$_{1-x}$Mn$_x$As]. Finally, we mention that the magnetic interaction
Hamiltonian defined by Eq.~(\ref{ham}) is sometimes referred to as the
{\it s-d} (or {\it s-f}) exchange Hamiltonian \cite{Kasuya} or the
Zener model \cite{sds22} in the literature although it was originally
introduced \cite{sds23} by Nabarro and Fr\"{o}hlich in a slightly
different context.  The physics we are interested in is how the local
exchange interaction defined in Eqs.~(\ref{ham}) and (\ref{ham_full})
could lead to global ferromagnetic ordering of the impurity local
moments below a Curie temperature $T_c$. We mention that we choose
$S=5/2$ and $s=1/2$ in all our numerical calculations below.

Unfortunately none of the parameters $J$, $n_{\textrm{c}}$,
$n_{\textrm{i}}$ of our model is directly experimentally measurable.
This is why we have emphasized throughout this work qualitative
behavior in temperature-dependent magnetization as a function of the
system parameters. The carrier density $n_{\textrm{c}}$ is hard to
measure even in metallic ferromagnetic materials because of the
problems associated with anomalous Hall effect (and the situation is
obviously worse in strongly insulting systems). The local moment
density $n_{\textrm{i}}$ is unknown because only a fraction of the
incorporated Mn atoms are magnetically active due to the invariable
presence of Mn interstitials and other possible defects in the system.
What is known about Ga$_{1-x}$Mn$_x$As is that it is a heavily
compensated system in the sense that the density of holes, is much
less than the density of Mn, typically
$n_{\textrm{c}}/n_{\textrm{i}}\sim 0.1$.  Much of this compensation
most likely arises from various defects invariably present in
low-temperature MBE grown Ga$_{1-x}$Mn$_x$As. The two most ``harmful''
defects in this respect are Mn interstitials and As antisites. Both of
these defects act as effective double donors, producing two
electrons each. Thus the holes produced by the magnetically and
electrically active ``desirable'' Mn$^{2+}$ ions (sitting at
substitutional cation sites), with each substitutional Mn ion
producing one hole, could be heavily compensated by the defects
leading to the existing situation in Ga$_{1-x}$Mn$_x$As where
$n_{\textrm{c}}/n_{\textrm{i}}\ll 1$.

All aspects of Kondo physics are completely negligible for the current
problem. Kondo physics is relevant only in the complementary regime of
high carrier density ($n_{\textrm{c}} / n_{\textrm{i}}\gg 1$), where a
paramagnetic carrier ground state entails as the impurity spin is
quenched by the free carriers.  For itinerant delocalized carriers
({\it i.\ e.}, metallic DMS), the indirect exchange coupling between the
local moments (induced by carrier spin polarization) is precisely the
RKKY interaction.  The relevance (or perhaps even, the dominance) of
RKKY physics (leading to a magnetic ground state) over the Kondo
physics in the low carrier density limit of the Kondo lattice system
has occasionally been mentioned in the Kondo-effect literature.\cite{R2}

The crucial element of RKKY physics playing a key role in DMS
ferromagnetism is the relatively low values of carrier density in
these materials leading to $k_F \sqrt[3]{n_{\textrm{i}}} \ll 1$ (where
$k_F$ is the Fermi wavevector associated with the carrier density
$n_{\textrm{c}}$ and $\sqrt[3]{n_{\textrm{i}}}$ is the characteristic
inter-impurity separation) so that the RKKY interaction is essentially
always ferromagnetic, and the RKKY spin-glass type behavior
predominant in disordered magnetic metallic systems may not arise here
since the interaction is mostly ferromagnetic avoiding effects of
frustration.

In principle one could start from the general Hamiltonian
(\ref{ham_full}), and try to develop a theory for carrier localization
and magnetism on an equal footing.  Such an ambitious attempt would be
essentially futile, due to the enormous complexity of the problem,
which would require both disorder and exchange interaction to be
treated non-perturbatively.  We therefore adopt the reasonable
empirical approach of building into our basic model the metallic
(itinerant carriers) or the insulating (localized carriers) nature of
the system, and develop separate complementary theoretical approaches
for the two situations in order to compare and contrast the nature of
DMS ferromagnetism in metallic and insulating systems, both with the
local exchange magnetic Hamiltonian (\ref{ham}) coupling the impurity
moments to carrier spins.

We use complementary theoretical approaches (degenerate- and
non-degenerate-carrier Weiss static mean-field theory, dynamical mean
field theory (DMFT), and percolation theory) in this paper; two of
which apply to the limiting cases of extended ``metallic'' system
(degenerate-carrier mean-field theory and DMFT) and the other two to
the strongly localized ``insulating'' system (non-degenerate-carrier
mean-field theory and percolation theory).  In principle, DMFT can
interpolate smoothly between the systems with extended and localized
carriers, but we have neglected localization effects in our DMFT
calculations carried out so far.  Our results for temperature- and
magnetic-field-dependent magnetization $M(T,B)$ exhibit qualitative
behavior very similar to that seen in experiments which, given the
minimal nature of our theoretical model, is all we can expect of the
theory. In our view, at this early stage of the development of DMS
materials and their physical understanding (and given the metastable
and fragile complex nature of the DMS systems), where all the
materials details (\emph{e.\ g.}\ defects and disorder of many
possible types) qualitatively affecting the system magnetization may
not even be known at the present time, it is premature to demand (or
impose, for example, by tuning the parameters of the theoretical
model) quantitative agreement between theory and experiment.  In this
respect we strongly disagree with the statements asserting that DMS
ferromagnetism is a well-understood problem based on a model of
free valence band holes interacting with Mn local moments. The
excellent qualitative agreement between our calculated magnetization
results in different theoretical approaches and the experimental
magnetization data in the existing literature provides useful insight
into the possible magnetic mechanisms underlying ``metallic'' and
``insulating'' DMS systems.

\subsection{The Static Mean-Field Theory}\label{sec:2a}

The basic idea underlying the static mean-field theory, as applied to
ferromagnetism in DMS, is to represent action of all impurity/hole
spins upon a given impurity/hole spin as an effective ``mean field,''
whose value is determined by the average values of the spins acting
upon this given spin. The resulting equations for the spins of
impurities and holes are to be solved self-consistently, finally
yielding the equilibrium magnetization at a given temperature.

The difference between the mean-field theory considered in this
Section and the canonical Weiss mean-field model arises from the
existence of two interacting species of spins, those of holes and of
impurities in a DMS system.  As a result, we have two effective
fields -- one determined by the average value of a hole spin and
the other determined by the average value of an impurity spin. In the
framework of Hamiltonian~(\ref{ham_full}), the effective field acting
upon holes has contributions coming from magnetic impurities and
from the external magnetic field $\textbf{B}$,
\begin{equation}
\label{Beffh}
B^{(\textrm{c})}_{\textrm{eff}}= \frac{1}{g_{\textrm{c}}\mu_B}
\left(
Ja_0^3 n_{\textrm{i}}\langle S_z \rangle
\right)
+B
\;,
\end{equation}
where the direction of the $z$ axis is chosen to coincide with the
direction of applied magnetic field $\mathbf{B}$, or, in the case of
$B=0$, with the direction of spontaneous magnetization of impurities.
The effective field acting upon impurities is a sum of contributions
from holes and the external magnetic field. Relative simplicity of the
static mean-field theory allows us to account also for possible direct
antiferromagnetic interaction between magnetic impurities by adding
the following term to the Hamiltonian
\begin{equation}
H_{\textrm{AF}}  = \sum_{jk} J^\textrm{{AF}}_{jk}
\left(\mathbf{S}_j \cdot \mathbf{S}_k\right)\;,
\label{H_AF}
\end{equation}
which yields one more contribution to the effective field acting upon
magnetic impurities:
\begin{equation}
\label{Beffi}
B^{(\textrm{i})}_{\textrm{eff}}= \frac{1}{g_{\textrm{i}}\mu_B}
\left(
z^{\textrm{AF}}J^{\textrm{AF}}\langle S_z \rangle +
Ja_0^3 n_{\textrm{c}}\langle s_z \rangle
\right)
+B \;,
\end{equation}
where $z^{\textrm{AF}}$ is the effective number of surrounding
impurities a given impurity interacts with.

The response of the impurity spin to this effective field
$B^{(\textrm{i})}_{\textrm{eff}}$ is given by
\begin{equation}
\label{siofB}
\left\langle S_z \right\rangle = 
S\mathcal{B}_S
\left(\frac{g_{\textrm{i}}\mu_B B^{(\textrm{i})}_{\textrm{eff}}}{k_B T}\right)
\end{equation}
where 
\begin{equation}
\label{brilloin}
\mathcal{B}_s(x) \equiv
\frac{2s+1}{2s}\coth\frac{2s+1}{2s}x-\frac{1}{2s}\coth\frac{1}{2s}x
\end{equation}
is the Brillouin function.  The magnetic response of hole spins to
effective field $B^{(\textrm{c})}_{\textrm{eff}}$ produced by
impurities strongly depends on whether the hole gas is degenerate or
not. The two complementary cases of non-degenerate and degenerate
holes will be considered below in Secs.~\ref{sec:2a1} and
\ref{sec:2a2} respectively. We mention that the non-degenerate
(degenerate) situation applies primarily to the insulating
(``metallic'') DMS systems.

\subsubsection{Non-degenerate holes}\label{sec:2a1}

The case of non-degenerate holes, when the hole spin distribution is
not affected by the Pauli exclusion principle, corresponds to two
physical situations. The first one occurs when the holes are localized
with strong on-site repulsion, so there is only one hole at each
localization center. This scenario is relevant to DMS with localized
carriers, which will also be considered in Sec.~\ref{sec:2c} using the
percolation theory. The second situation takes place in a gas of
delocalized holes when the temperature is higher than the Fermi
energy.

When the Pauli exclusion principle plays no role in the spin
distribution of electrons, the latter is determined by Boltzmann
statistics, and the average hole spin, as determined by the effective
mean-field $B^{(\textrm{c})}_{\textrm{eff}}$ is given by
\begin{equation}
\label{shofB}
\left\langle s_z \right\rangle = 
S\mathcal{B}_s
\left(
\frac{g_{\textrm{c}}\mu_B B^{(\textrm{c})}_{\textrm{eff}}}{k_B T}
\right)\;,
\end{equation}
similarly to Eq.~(\ref{siofB}), with $g_{\textrm{c}}$ being the hole
$g$-factor.

Combining Eqs.~(\ref{Beffh}), (\ref{Beffi}), (\ref{siofB}), and
(\ref{shofB}), we obtain the self-consistent equation for
$\left\langle S_z \right\rangle$:
\
\begin{widetext}
\begin{eqnarray}
\label{sieq}
\frac{\left\langle S_z \right\rangle}{S}&=&\mathcal{B}_S\!
\left[-3\frac{T_{c0}}{T}\sqrt{\frac{n_{\textrm{c}}}{n_{\textrm{i}}}}
\sqrt{\frac{sS}{(s+1)(S+1)}}
\mathcal{B}_s\!\left(
-3\frac{T_{c0}}{T}\sqrt{\frac{n_{\textrm{i}}}{n_{\textrm{c}}}}
\sqrt{\frac{sS}{(s+1)(S+1)}}\frac{\left\langle S_z \right\rangle}{S}
+\frac{g_{\textrm{c}}\mu_B B}{k_B T}
\right)\right.\nonumber\\
&&\qquad+
\left.\vphantom{\sqrt{\frac{sS}{(s+1)(S+1)}}}
\frac{6ST_1}{(S+1)T}\frac{\left\langle S_z \right\rangle}{S}
-\frac{g_{\textrm{i}}\mu_B B}{k_B T}
\right],
\end{eqnarray}
\end{widetext}
where
\begin{equation}
\label{Tc0}
k_B T_{c0}=\frac13 Ja_0^3 \sqrt{n_{\textrm{c}}n_{\textrm{i}}}
\sqrt{S(S+1)s(s+1)}\;,
\end{equation}
\begin{equation}
\label{T1}
k_B T_1 = \frac16 S(S+1) z^{\textrm{AF}} J^{\textrm{AF}}\;,
\end{equation}
and $J$ is assumed to be negative (antiferromagnetic interaction
between impurities and holes).  Similarly to the text-book mean-field
theory, Eq.~(\ref{sieq}) has a non-trivial solution in the absence of
the external magnetic field only if temperature $T$ is below a certain
value, which is the ferromagnetic transition temperature $T_c$. Using
the expansion for the Brillouin function
\begin{equation}
\label{brillexp}
\left.\mathcal{B}_s(x)\right|_{x\ll 1} \approx \frac{s+1}{3s} x + O(x^3)\;,
\end{equation}
we arrive at the following expression:
\begin{equation}
\label{Tcn}
T_c^{\textrm{(n)}}=\sqrt{T_{c0}^2+T_1^2}-T_1\;.
\end{equation}

\begin{figure}
\includegraphics[width=3in]{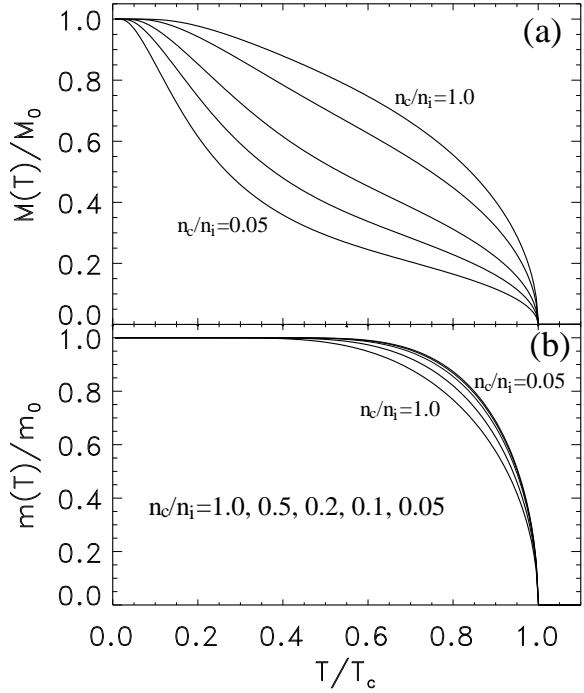}
\vspace{0.5cm}
\caption{ (a) Dopant magnetization $M/M_0\equiv \langle S_z \rangle
  /S$ and (b) hole magnetization $m/m_0\equiv \langle s_z \rangle /s$
  with $S=5/2$ and $s=1/2$ from the non-degenerate-hole model for
  various density ratio ($n_{\textrm{c}}/n_{\textrm{i}} =$ 1.0, 0.5,
  0.2, 0.1, 0.05).}
\label{mft_fig1}
\end{figure}

In the absence of antiferromagnetic interaction between the
impurities, $T_1=0$ so the ferromagnetic transition temperature
equals $T_{c0}$. Equation (\ref{sieq}) in the absence of the external
magnetic field and at $J^{\textrm{AF}}=0$ reduces to
\begin{eqnarray}
\label{sieqreduced}
\!\!\!\!\!\!\!\!
\frac{\left\langle S_z \right\rangle}{S}\!\! &=&\!\! \mathcal{B}_S\!
\left[3\frac{T_{c0}}{T}\sqrt{\frac{n_{\textrm{c}}}{n_{\textrm{i}}}}
\sqrt{\frac{sS}{(s+1)(S+1)}}\right.\nonumber\\
&&\quad\!\!\times\left.\!\!
\mathcal{B}_s\!\left(
3\frac{T_{c0}}{T}\sqrt{\frac{n_{\textrm{i}}}{n_{\textrm{c}}}}
\sqrt{\frac{sS}{(s+1)(S+1)}}\frac{\left\langle S_z \right\rangle}{S}
\right)
\right]\!,
\end{eqnarray}
the average hole spin is still given by Eqs.~(\ref{shofB}) and
(\ref{Beffh}) with $B=0$. The solution 
of Eq.~(\ref{sieqreduced}) can be found numerically.  The resulting
magnetization curves for impurities and holes for several values of
$n_{\textrm{c}}/n_{\textrm{i}}$ are shown in Fig.~\ref{mft_fig1}.
Note that specific values of $J$, $n_{\textrm{i}}$, and $n_{\textrm{c}}$ are not of any
relevance here -- only the ratio $n_{\textrm{c}}/n_{\textrm{i}}$ is
the important tuning parameter in determining magnetization as a
function of $T/T_c$.  The most important salient feature of
Fig.~\ref{mft_fig1} is the highly ``non-mean-field-like'' concave
magnetization behavior for low values ($\le 0.2$) of
$n_{\textrm{c}}/n_{\textrm{i}}$. The reason for such behavior is that
if we have one hole per many impurities, the effective field
$B^{(\textrm{c})}_{\textrm{eff}}$ acting on holes is much stronger
than its counterpart $B^{(\textrm{i})}_{\textrm{eff}}$ acting on
impurities. As a result, the hole magnetization grows as
\begin{displaymath}
\frac{\left\langle s_z \right\rangle}{s} \sim \sqrt{\frac{T_{c0}^{\textrm{(n)}}-T}{T}}
\end{displaymath}
for $T\lesssim T_{c0}$ and reaches unity at some temperature of the
order of $T_{c0}$, while the impurity magnetization is still much less
than unity. As the temperature is getting lower, the impurity
magnetization grows as
\begin{displaymath}
\frac{\left\langle S_z \right\rangle}{S}\approx
\mathcal{B}_S\!
\left(3\frac{T_{c0}^{\textrm{(n)}}}{T}\sqrt{\frac{n_{\textrm{c}}}{n_{\textrm{i}}}}
\sqrt{\frac{sS}{(s+1)(S+1)}}
\right)
\end{displaymath}
and approaches unity only at
\begin{displaymath}
T\sim T^{\textrm{(n)}}_{c0}\sqrt{\frac{n_{\textrm{c}}}{n_{\textrm{i}}}}\ll T_{c0}^{\textrm{(n)}}\;.
\end{displaymath}

\begin{figure}
\includegraphics[width=3in]{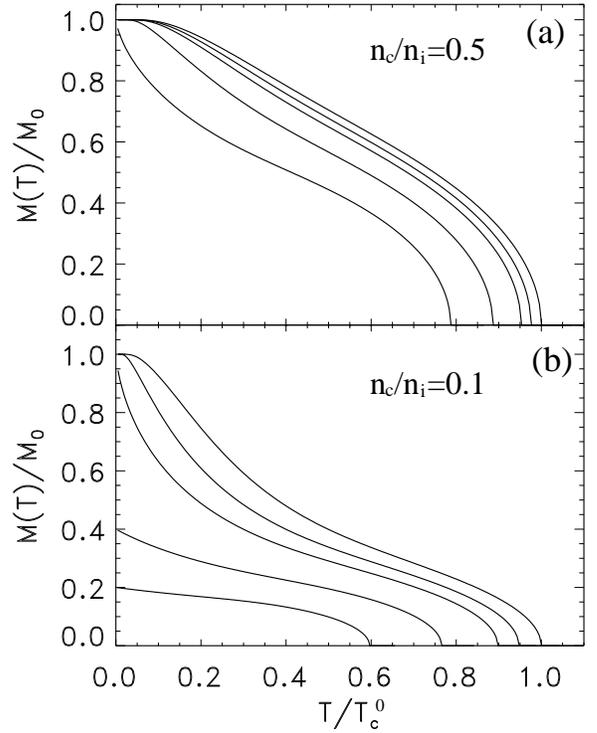}
\caption{Dopant magnetization with both $J^{\textrm{AF}}$ and $J$
  from the non-degenerate-hole model with $S=5/2$ and $s=1/2$ for
  various coupling constant ratio ($z^{\textrm{AF}}J^{\textrm{AF}}/J =$ 0.0, 0.5, 1.0,
  2.5, 5.0$\times 10^{-3}$, from the top).  (a) for density ratio
  $n_{\textrm{c}}/n_{\textrm{i}} = 0.5$ and (b)
  $n_{\textrm{c}}/n_{\textrm{i}}=0.1$. }
\label{mft_fig2}
\end{figure}


\begin{figure}
\includegraphics[width=3in]{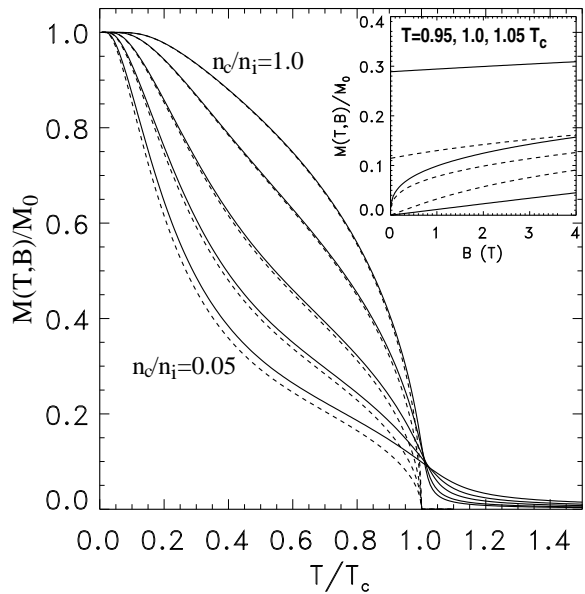}
\caption{External magnetic field and temperature
  dependence of dopant magnetization for the non-degenerate-hole model for
  various density ratio ($n_{\textrm{c}}/n_{\textrm{i}} =$ 1.0, 0.5,
  0.2, 0.1, 0.05, from the top). Solid (dashed) lines indicate the
  results for a fixed external magnetic field $B=2.6T$ ($0T$). The
parameters of Fig. 1 are used.  In the inset the magnetization curves as a
function of magnetic field are given for fixed temperatures ($T=$0.95,
1.0, 1.05$T_c$, from the top). Solid (dashed) lines represent the
results for $n_{\textrm{c}}/n_{\textrm{i}}=1.0$
($n_{\textrm{c}}/n_{\textrm{i}}=0.1$).  }
\label{mft_fig3}
\end{figure}

Finite antiferromagnetic coupling $J^{\textrm{AF}}$ suppresses $T_c$, as one
can easily see from Eq.~(\ref{Tc}). In Fig. \ref{mft_fig2} 
we show the dopant magnetization, as given by the (numerical) solution
of Eq.~(\ref{sieq}) for $B=0$,
for various values of $T_1/T_{c0}$ for two
typical values of $n_{\textrm{c}}/n_{\textrm{i}}=0.5$ and $0.1$.
 
In discussing the results shown in Fig.~\ref{mft_fig2} we first note
that the actual direct Mn-Mn antiferromagnetic coupling
$J^{\textrm{AF}}$ is expected to be very small in Ga$_{1-x}$Mn$_{x}$As
for low values of $x$, {\it i.e.} for relatively large Mn-Mn
separation, since the direct antiferromagnetic coupling falls off
exponentially with inter-atomic distance. For larger values of $x$,
however, effects of antiferromagnetic Mn-Mn coupling may very well be
important in determining the DMS magnetic behavior. In general, finite
$J^{\textrm{AF}}$ suppresses both $T_c$ and $M(T)$ as one would
expect. In particular, the zero-temperature magnetization
$M(T\rightarrow 0)$ could be strongly suppressed far below the
saturation magnetization by the finite AF coupling, particularly for
lower hole densities,
\begin{equation}
\label{satmagaf}
\left.\frac{\left\langle S_z \right\rangle}{S}\right|_{T\to0}=
\min\left\{
1\;,\;
\frac12 \frac{T^{\textrm{(n)}}_{c0}}{T_1}
\sqrt{\frac{n_{\textrm{c}}}{n_{\textrm{i}}}}
\sqrt{\frac{S+1}{S}\frac{s}{s+1}}
\right\}\;.
\end{equation}

An interesting feature of Fig.~\ref{mft_fig2} is that at low carrier
densities larger values of $J^{\textrm{AF}}$
 may actually restore (albeit with strongly suppressed value of saturation
magnetization) the convex $M(T)$ shape (\emph{e.\ g.}\ the
$z^{\textrm{AF}}|J^{\textrm{AF}}|/J=5.0\times 10^{-3}$
curve for $n_{\textrm{c}}/n_{\textrm{i}}=0.1$ in Fig.~\ref{mft_fig2}(b)).
For higher hole densities, however, $M(T)$ becomes more concave for
$J^{\textrm{AF}} \neq 0$.

The magnetic susceptibility of the system is essentially that of
impurities (since $n_{\textrm{c}}/n_{\textrm{i}}\ll 1$ and $S>s$), and
above the critical temperature is given by
\begin{eqnarray}
\chi(T)&\equiv& n_{\textrm{i}}
\frac{\partial(g_{\textrm{i}}\mu_B \langle S_z \rangle)}{\partial B}
\nonumber\\
& =& 
\frac{\chi_0 + 
\displaystyle\frac{\left(T^{\textrm{(n)}}_{c0}\right)^2}{T^2 Ja_0^3}
g_{\textrm{i}}g_{\textrm{c}}\mu_B^2
}
{\left(1-\displaystyle\frac{T_c^{\textrm{(n)}}}{T}\right)
\left(1+\displaystyle\frac{T^*}{T}\right)},
\label{chinondeg}
\end{eqnarray}
where $T^* \equiv T_1 + \sqrt{T_1^2 + T_{c0}^2}$, and $\chi_0 =
n_{\textrm{i}} (g_{\textrm{i}}\mu_B)^2 S(S+1)/3T$ is the paramagnetic
susceptibility of bare impurities.

Finally, we show in Fig.~\ref{mft_fig3} the effect of an external
magnetic field by showing $M(T,B)$ for fixed parameter values {\it J,
  S, s}, and for various density ratio
$n_{\textrm{c}}/n_{\textrm{i}}$.  In the main figure we show $M(T,B)$
as a function of temperature for a magnetic field value, $B=2.6T$
(solid lines), and in the inset we show $M(T,B)$ as a function of the
external magnetic field for different temperatures.  
In inset we show $M(B,T_c) \propto
B^{1/3}$ as expected in the mean-field theory.

\subsubsection{Degenerate hole gas} \label{sec:2a2}

For metallic DMS systems where ferromagnetic $T_c$ is optimum (at
least for Ga$_{1-x}$Mn$_x$As), the carrier system is typically
delocalized, with the Fermi energy substantially exceeding the
ferromagnetic transition temperature. In such systems, the formalism
of Sec.~\ref{sec:2a1}, developed for the non-degenerate hole system is
not applicable anymore, and we have to use the generic expression
\begin{eqnarray}
\label{degspin}
\frac{\langle s_z\rangle}{s} &= &
\frac{1}{n_{\textrm{c}}}\frac{1}{2}
\int d\varepsilon\,f(\epsilon)
\left[D(\varepsilon + g_{\textrm{c}}\mu_B s B_{\textrm{eff}}^{\textrm{(c)}})
\right.
\nonumber\\
&&\qquad\qquad\qquad\left.
-D(\varepsilon - g_{\textrm{c}}\mu_B s B_{\textrm{eff}}^{\textrm{(c)}})
\right]
\end{eqnarray}
for the hole polarization, where $D(\varepsilon) $ is the hole density
of states.

If the effective field $B_{\textrm{eff}}^{\textrm{(c)}}$ acting on
carriers is weak, we can expand the density of states up to the first
order in the effective field to obtain
\begin{equation}
\frac{\langle s_z\rangle}{s} = 
\frac{1}{n_{\textrm{c}}}
g_{\textrm{c}}\mu_B s B_{\textrm{eff}}^{\textrm{(c)}}
\int d\varepsilon\,f(\epsilon)
D'(\varepsilon)\;.
\label{degspin1}
\end{equation}
For very low temperatures, $k_B T \ll E_F$ (in general, this condition
is satisfied in metallic Ga$_{1-x}$Mn$_x$As systems where typically
$k_B T_c/E_F < 0.1$ noting that both $T_c$ and $E_F$ decrease with
lowering the hole density $n_{\textrm{c}}$), we get
\begin{equation}
\frac{\langle s_z\rangle}{s} = 
\frac{1}{n_{\textrm{c}}}
g_{\textrm{c}}\mu_B s B_{\textrm{eff}}^{\textrm{(c)}}
D'(E_F)\;.
\label{degspin2}
\end{equation}
Linearizing Eq.~(\ref{siofB}) and using Eqs.~(\ref{Beffh}) and
(\ref{Beffi}), we arrive at
\begin{equation}
T^{\textrm{(d)}}_c = T^{\textrm{(d)}}_{c0} - 2 T_1\;,
\label{tc02}
\end{equation}
where
\begin{equation}
\label{tcdeg}
T^{\textrm{(d)}}_{c0}= \frac{S(S+1)}{3}s^2
(Ja_0^3)^2 D(E_F) n_{\textrm{i}}
\end{equation}
and $T_1$ is given by Eq.~(\ref{T1}). 

In the absence of antiferromagnetic interaction between magnetic
impurities, $J^{AF}=0$, the ferromagnetic transition temperature
equals $T_{c0}$. As one can see from Eq.~(\ref{tcdeg}), $T_{c0}
\propto J^2 n_{\textrm{c}}^{1/3}$ in the Weiss mean-field theory for
degenerate holes (since $D(E_F)\propto n_{\textrm{c}}^{1/3}$ for a
three-dimensional degenerate electron gas), in contrast to the
non-degenerate case considered in Sec.~\ref{sec:2a1}, where $T_{c0}
\propto J n_{\textrm{c}}^{1/2}$.

When the hole density is very small, the hole Fermi energy $E_F$ may
be comparable to the effective magnetic energy $g_{\textrm{c}}\mu_B s
B_{\textrm{eff}}^{\textrm{(c)}}$.  In this case we can not expand
Eq.~(\ref{degspin}) with respect to the effective field
$B_{\textrm{eff}}^{\textrm{(c)}}$, and the hole magnetization must be
obtained by directly integrating Eq.~(\ref{degspin1}).

Before presenting our numerical results for the calculated
temperature-dependent DMS magnetization for the degenerate-carrier
mean-field theory, we mention that equation (\ref{tc02}) for the
ferromagnetic transition temperature in the Zener model was first
derived in Ref.~\onlinecite{AG} more than forty years ago and has
recently been rediscovered \cite{Dietl,Macdonald} in the context of
DMS ferromagnetism.

\begin{figure}
\includegraphics[width=3in]{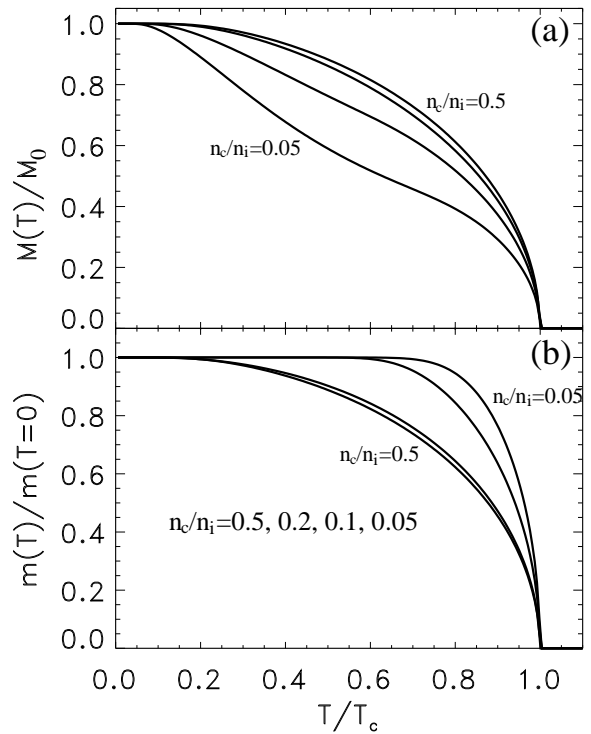}
\caption{ (a) Dopant magnetizations from the 
  degenerate-hole model for various value of the density ratio
  ($n_{\textrm{c}}/n_{\textrm{i}} =$ 0.5, 0.2, 0.1, 0.05, from the
  top). Here we use the fixed parameter values: $S=5/2$, $s=1/2$,
  $n_{\textrm{i}}=10^{21}$ cm$^{-3}$, $m=0.5m_e$, $x=0.05$, and coupling
  constant $J = 3.0$ eV.  (b) Hole magnetization with the same
  parameters as in (a). Note $m(T=0)$ is the magnetization at $T=0$,
  which may not be equal to $m_0 \equiv g_c\mu_B s n_{\textrm{c}}$.}
\label{mft_fig4}
\end{figure}

In Fig.~\ref{mft_fig4} we show our calculated impurity and carrier
magnetization for the degenerate-carrier Weiss mean-field theory using
the same parameters (typical for Ga$_{1-x}$Mn$_x$As) as for the
corresponding non-degenerate case shown in Fig.~\ref{mft_fig1}. In
general, except for the lowest hole density with
$n_{\textrm{i}}/n_{\textrm{c}}=0.05$, the impurity magnetization is
convex, but looks quite different from the classic text book Weiss
form except perhaps at very high carrier densities when
$n_{\textrm{i}}/n_{\textrm{c}} \approx 1$. In particular, for the
realistic value of $n_{\textrm{c}}/n_{\textrm{i}}=0.1$ (which is
thought to apply to many Ga$_{1-x}$Mn$_x$As samples), $M(T)$ in
Fig.~\ref{mft_fig4} has the non-mean-field-like straight line shape at
lower temperatures. For higher values of hole density,
$n_{\textrm{c}}/n_{\textrm{i}}=$0.2 and 0.5, the magnetization is much
closer to the classical Brillouin shape whereas for lower carrier
densities (\emph{e.\ g.}, $n_{\textrm{c}}/n_{\textrm{i}}=0.05$) even
this metallic DMS system starts exhibiting the concave magnetization
curve typical of the insulating DMS systems discussed in the last
sub-section (\ref{sec:2a1}) of this article.

The origin of the concave shape of the magnetization curve is exactly
the same as for the non-degenerate case discussed in
Sec.~\ref{sec:2a1}, namely, the magnetization of holes saturates at
$T\lesssim T_{c0}$, much earlier than magnetization of impurities,
which, upon saturation of the hole magnetization grows as given by
Eq.~(\ref{sieqreduced}), which represents a concave curve provided
$n_{\textrm{c}}/n_{\textrm{i}}\ll 1$. Thus, the concave magnetization
behavior may be generic to the low carrier density limit of DMS
ferromagnets, independent of whether they are metallic or insulating
although the highly concave magnetization curves of
Fig.~\ref{mft_fig1}(a) are clearly much more typical of insulating DMS
systems than the metallic ones. The carrier magnetization results
presented in Fig.~\ref{mft_fig4}(b) are very similar to textbook
convex Weiss magnetization behavior. We note that the
degenerate-carrier mean-field theory results shown in
Fig.~\ref{mft_fig4}(a) are qualitatively very similar to the
experimentally measured temperature dependent magnetization data in
metallic Ga$_{1-x}$Mn$_x$As systems.  In particular, recent annealing
experiments, where annealing leads to better metallicity (\emph{e.\ 
  g.}, higher conductivity) by virtue of increasing the hole density,
show qualitative trends strikingly similar to the results of Fig.
\ref{mft_fig4}(a).

\begin{figure}
\includegraphics[width=3in]{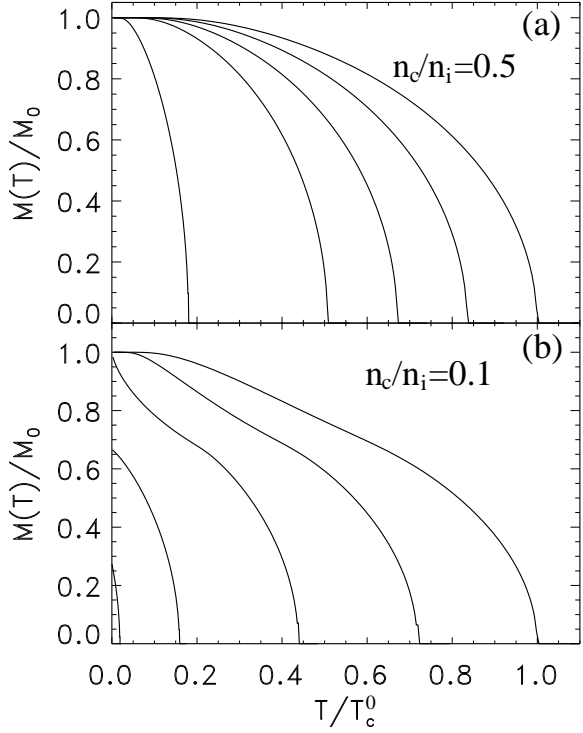}
\caption{Dopant magnetization with both $J^{\textrm{AF}}\neq 0$ 
  from the degenerate-hole model for various values of the coupling
  constant ratio (a) $z^{\textrm{AF}}J^{\textrm{AF}}/J=0.0$, 0.5,
  1.0, 1.5, 2.5$\times 10^{-3}$ (from the top) and for density ratio
  $n_{\textrm{c}}/n_{\textrm{i}} = 0.5$, (b)
  $z^{\textrm{AF}}J^{\textrm{AF}}/J=0.0$, 0.5, 1.0, 1.5, 1.75$\times
  10^{-3}$ (from the top) and $n_{\textrm{c}}/n_{\textrm{i}}=0.1$.
  The parameter values of Fig. \ref{mft_fig4} are used.  }
\label{mft_fig5}
\end{figure}

Next we include the direct antiferromagnetic coupling (\ref{H_AF})
between the local moments in our consideration. With
$J^{\textrm{AF}}\neq 0$, the critical temperature is given by
Eq.~(\ref{tc02}). The calculated impurity magnetization in the
presence of the antiferromagnetic coupling is shown in
Fig.~\ref{mft_fig5}.  Comparing with the non-degenerate-hole model we
find that the degenerate hole case is more quantitatively sensitive to
the antiferromagnetic coupling although the qualitative effect of a
finite $J^{\textrm{AF}}$ in both metallic and insulating DMS materials is
basically the same, namely suppression of the transition temperature
and the magnetization.

Similar to the results shown for the insulating systems, we present in
Fig. \ref{mft_fig5} calculated impurity magnetization for two values
of the ratio $n_{\textrm{c}}/n_{\textrm{i}}=0.5$
[Fig.~\ref{mft_fig5}(a)] and 0.1 [Fig.~\ref{mft_fig5}(b)].

\begin{figure}
\includegraphics[width=3in]{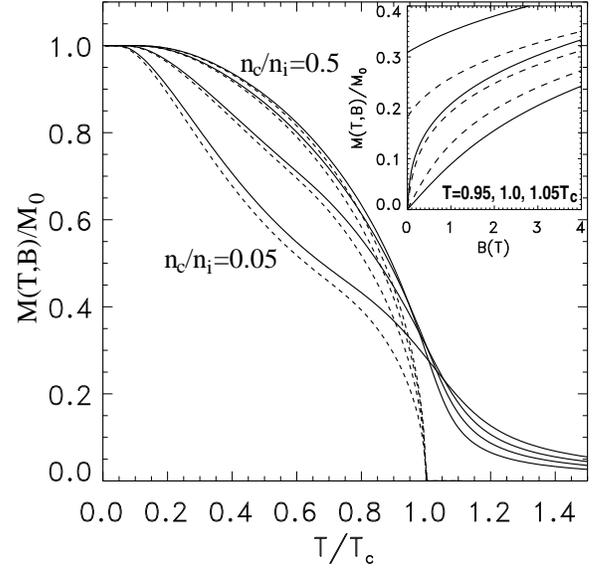}
\caption{ 
  External magnetic field and temperature dependence of dopant
  magnetization from the degenerate-hole model for various values of
  the density ratio ($n_{\textrm{c}}/n_{\textrm{i}} =$ 0.5, 0.2, 0.1,
  0.05, from the top). Solid (dashed) lines indicate the results for a
  fixed external magnetic field $B=2.6T$ ($0T$). The parameters
  in Fig.  \ref{mft_fig4} are used.  In the inset the magnetization curves
  as a function of magnetic field are given for fixed temperatures
  ($T=$0.95, 1.0, 1.05$T_c^{\textrm{(d)}}$, from the top). Solid (dashed) lines
  represent the results for $n_{\textrm{c}}/n_{\textrm{i}} =0.5$
  ($n_{\textrm{c}}/n_{\textrm{i}} =0.05$).
\label{mft_fig6}}
\end{figure}

In Fig.~\ref{mft_fig6} the calculated external magnetic field and
temperature dependence of the dopant magnetization is shown for fixed
parameter values {\it J, S, s}, and for various density ratio
$n_{\textrm{c}}/n_{\textrm{i}}$.  In the main figure we show $M(T,B)$
as a function of temperature for a magnetic field value, $B=2.6T$
(solid lines), and in the inset we show $M(T,B)$ as a function of the
external field for fixed temperatures.  Our calculated $M(T,B)$
behavior is roughly qualitatively similar to experimental observations
in Ga$_{1-x}$Mn$_x$As systems in the ``metallic'' regime.

The magnetic susceptibility of the system is essentially that of
impurities, similarly to Sec.~\ref{sec:2a1}, Eq.~(\ref{chinondeg}),
and above the critical temperature is given by
\begin{equation}
\chi(T) =
\frac{\chi_0 + 
\displaystyle\frac{T_{c0}^{\textrm{(d)}}}{T Ja_0^3}
g_{\textrm{i}}g_{\textrm{c}}\mu_B^2
}
{1-\displaystyle\frac{T_c^{\textrm{(d)}}}{T}},
\end{equation}
where $\chi_0 = n_{\textrm{i}} (g_{\textrm{i}}\mu_B)^2 S(S+1)/3T$ is
the paramagnetic susceptibility of bare impurities.

The Weiss molecular mean-field theory results (for localized carriers
in Sec. \ref{sec:2a1} and for delocalized carriers in Sec.
\ref{sec:2a2}) presented above for DMS magnetization qualitatively
agree very well with the existing experimental data. In particular,
the basic trend of our results shown in Figs.
\ref{mft_fig1}-\ref{mft_fig6}, that the spontaneous magnetization is
strongly suppressed (perhaps even into a very unusual outwardly
concave shape) at low carrier densities and for more insulating
systems whereas at higher carrier densities and for more metallic
systems the magnetization has the usual convex textbook shape, is in
excellent agreement with experiment. Our degenerate-carrier mean-field
results (see Fig. 3(a)) also reproduce the almost linear
magnetization curves seen in GaMnAs for intermediate carrier
densities. But obviously one needs to go beyond the Weiss mean-field
theory for a deeper (and more quantitative) understanding of DMS
magnetization, if for no other reason than to validate (or to
ascertain the regime of validity of) the simple static mean-field
theory. At low carrier density the mean-field theory should work
better since on the average there are many Mn local moments between
any two holes, but typically the number of Mn atoms per hole is around
3-4 so the quantitative applicability of MFT is questionable.
Note that the hole-hole interaction neglected in our calculations may
be included in a crude approximate fashion by incorporating a Stoner
enhancement of the carrier susceptibility, though in general the
strength of this enhancement is unknown. 

In the next two sections we go beyond the Weiss static mean-field theory
and develop two more sophisticated approximation schemes to theoretically
study DMS magnetization.  These are DMFT (section \ref{sec:2b}) 
and the percolation theory (section \ref{sec:2c}).


\subsection{The Dynamical Mean-Field Theory (DMFT)}\label{sec:2b}

In this section we use a recently developed non-perturbative method,
the ``dynamical mean-field theory'' (DMFT),\cite{Kotliar} to calculate
magnetization for the minimal model (Eq.~(\ref{ham_full})) of dilute
magnetic semiconductors.  The DMFT has been recently applied to the
DMS system to calculate the magnetic transition temperature and the
optical conductivity \cite{sds6}.  DMFT is essentially a lattice
quantum version of the Weiss mean field theory where the appropriate
density of states (including impurity band formation) along with
temporal fluctuations are incorporated within an effective local field
theory.

We model the Ga$_{1-x}$Mn$_x$As system as a lattice of sites, which
are randomly nonmagnetic (with probability $1-x$) or magnetic (with
probability $x$), where $x$ now indicates the relative concentration
({\it i.e.} per Ga site) of active Mn local moments in GaMnAs. The
DMFT approximation amounts to assuming 
that the self energy is local or momentum independent, $\Sigma({\bf p},i\omega_n) 
\rightarrow \Sigma(i\omega_n)$, and then all of the relevant
physics may  be determined from the local (momentum-integrated) Green
function defined by
\begin{equation}
G_{\textrm{loc}}(i\omega_n) = a_0^3 \int \frac{d^3p}{(2\pi)^3} \frac{1}{
i\omega_n +\mu + h \sigma - \epsilon(p) 
-\Sigma_{\sigma}(i\omega_n)},
\end{equation}
where we have normalized the momentum integral to the volume of the
unit cell $a_0^3$, and $\mu$ is the chemical potential and $h$ the
external magnetic field.  $G_{\textrm{loc}}$ is in general a matrix in spin and
band indices and depends on whether one is considering a magnetic
($a)$ or non-magnetic ($b)$ site. Since $G_{\textrm{loc}}$ is a local function,
it is the solution of a local problem specified by a mean-field
function $g_0$, which is related to the partition function
$Z_{\textrm{loc}}=\int d\textbf{S} \exp(-S_{\textrm{loc}})$ with action
\begin{eqnarray}
S_{\textrm{loc}}&=&g_{0\alpha \beta }^{a}(\tau -\tau ^{\prime
})c_{u\alpha }^{+}(\tau 
)c_{u\beta }(\tau ^{\prime })\nonumber\\
&+&J \sum_{u\alpha \beta }\left(\textbf{S}\cdot\bm{\sigma}_{\alpha
  \beta}\right)
c_{u\alpha}^{+}(x)c_{u\beta }(x)\;, 
\end{eqnarray}
on the $a$ (magnetic) site and 
\begin{equation}
S_{\textrm{loc}}=g_{0\alpha \beta }^{b}(\tau -\tau ^{\prime })
c_{a\alpha }^{+}(\tau)c_{a\beta }(\tau ^{\prime }),
\end{equation}
on the non-magnetic ($b$) site.
The local Green function is calculated exactly as
\begin{equation}
G_{\textrm{loc}}(i\omega_n) = \left \langle \left (g_0^{-1} + J
{\textbf{S}} \cdot  
\bm{\sigma}_{\alpha\beta} \right )^{-1} \right \rangle,
\end{equation}
where the thermal average $\langle \dots \rangle $ is
taken with respect to the orientation of the local spin $\textbf{S}$.
The $a$-site mean-field function $g_{0}^{a}$ can be written as
$g_{0\alpha \beta }^{a}=a_{0}+a_{1}{\textbf{m}}\cdot
\bm{\sigma}_{\alpha \beta }$ with ${\textbf{m}}$ the
magnetization direction 
and $a_{1}$ vanishing in the paramagnetic state. It is specified by
the condition that the local Green function computed from
$Z_{\textrm{loc}}$, namely $\delta \ln Z_{\textrm{loc}}/\delta
g_{0}^{a}=\left( g_{0}^{a}-\Sigma \right) ^{-1}$ is identical to the
local Green function computed by performing the momentum integral
using the same self energy.

The form of the dispersion given in full Hamiltonian
Eq.~(\ref{ham_full}) applies only near the band edges.  It is
necessary for the method to impose a momentum cutoff, arising
physically from the carrier band-width.  We impose the cutoff by
assuming a semicircular density of states $D(\epsilon)=a_0^{3}\int
\frac{d^{3}p} {\left( 2\pi \right) ^{3}}\delta (\varepsilon
-\varepsilon _{pa}) = \sqrt{4t^2-\epsilon^2}/2\pi t$ with
$t=(2\pi)^{2/3}/ma_0^2$.  The parameter $t$ is chosen to correctly
reproduce the band edge density of states.  Other choices of upper
cutoff would lead to numerically similar results.  This choice of
cutoff corresponds to a Bethe lattice in infinite dimensions.
Other (perhaps more realistic) choices for the density of states would
give magnetization results qualitatively similar to our results since
the band edge density of states has the correct physical behavior in
our model. For this
$N(\epsilon)$ the self consistent equation for $g_{0}$ obeys the
equation
\begin{eqnarray}
{\bf g}_{0}^{a}(\omega )={\bf g}_{0}^{b}(\omega )&=&\omega +\mu
-xt^{2} \left \langle \left( {\bf g}_{0}^{a}(\omega )
+J{\bf S}\cdot {\bm \sigma }
_{\alpha \beta }\right) ^{-1} \right \rangle \nonumber\\
&-&(1-x)t^{2} \left \langle {\bf g}_{0}^{b}(\omega)^{-1} \right \rangle
,
\end{eqnarray}
where the angular brackets denote averages performed in the ensemble
defined by the appropriate $Z_{\textrm{loc}}$.

Within this approximation the normalized magnetization
$M(T)$ of the local moments is given by
\begin{equation}
M(T) = \int (d{\bf S}) \cdot \hat {\bf S}
\frac{\exp(-S_{\textrm{loc}})}{Z_{\textrm{loc}}}. 
\end{equation}
As the temperature is increased, the spins disorder and eventually the
magnetic transition temperature is reached. Above this temperature,
$g_{0}$ is spin-independent.  By linearizing the equation in the
magnetic part of $g_0$ with respect to $a_1$ we may obtain the
ferromagnetic transition temperature $T_c$.  The details on the
calculation of $T_c$ are
given in Ref.~\onlinecite{sds6}.  The critical temperature $T_c$
and magnetization of the local moment depend crucially on $J/t$, $x$,
and carrier density $n_{\textrm{c}}$.
Note that in the DMFT calculation it is more convenient to express our
results in terms of $n$, the relative concentration of active local
moments rather than $n_i$, the absolute impurity density. 

\begin{figure}
\includegraphics[width=3.0in]{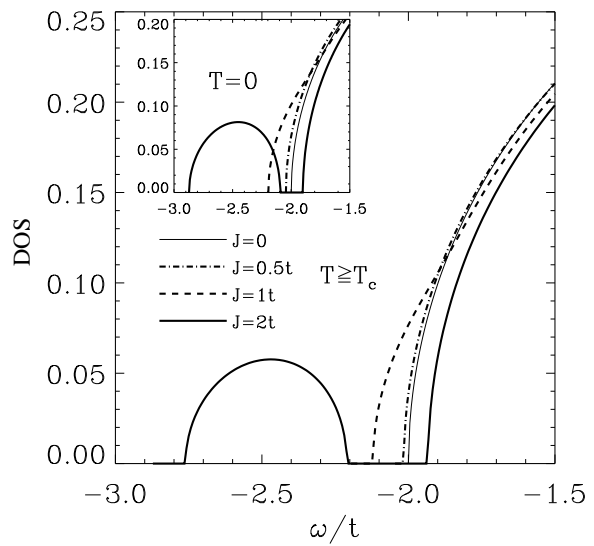}
\caption{
\label{DMFT_fig7}
The calculated density of states for dynamical mean 
field calculations applied to the semicircular model.
Shown is the evolution of majority spin
DOS for various carrier-spin coupling $J$ in the disordered 
spin $T>T_c$ state. Inset shows the DOS for the ordered ferromagnetic
state at $T=0$. For the large coupling constant $J>J_c$ we find a well
separated impurity band below the main band.}
\end{figure}

\begin{figure}
\includegraphics[width=2.5in]{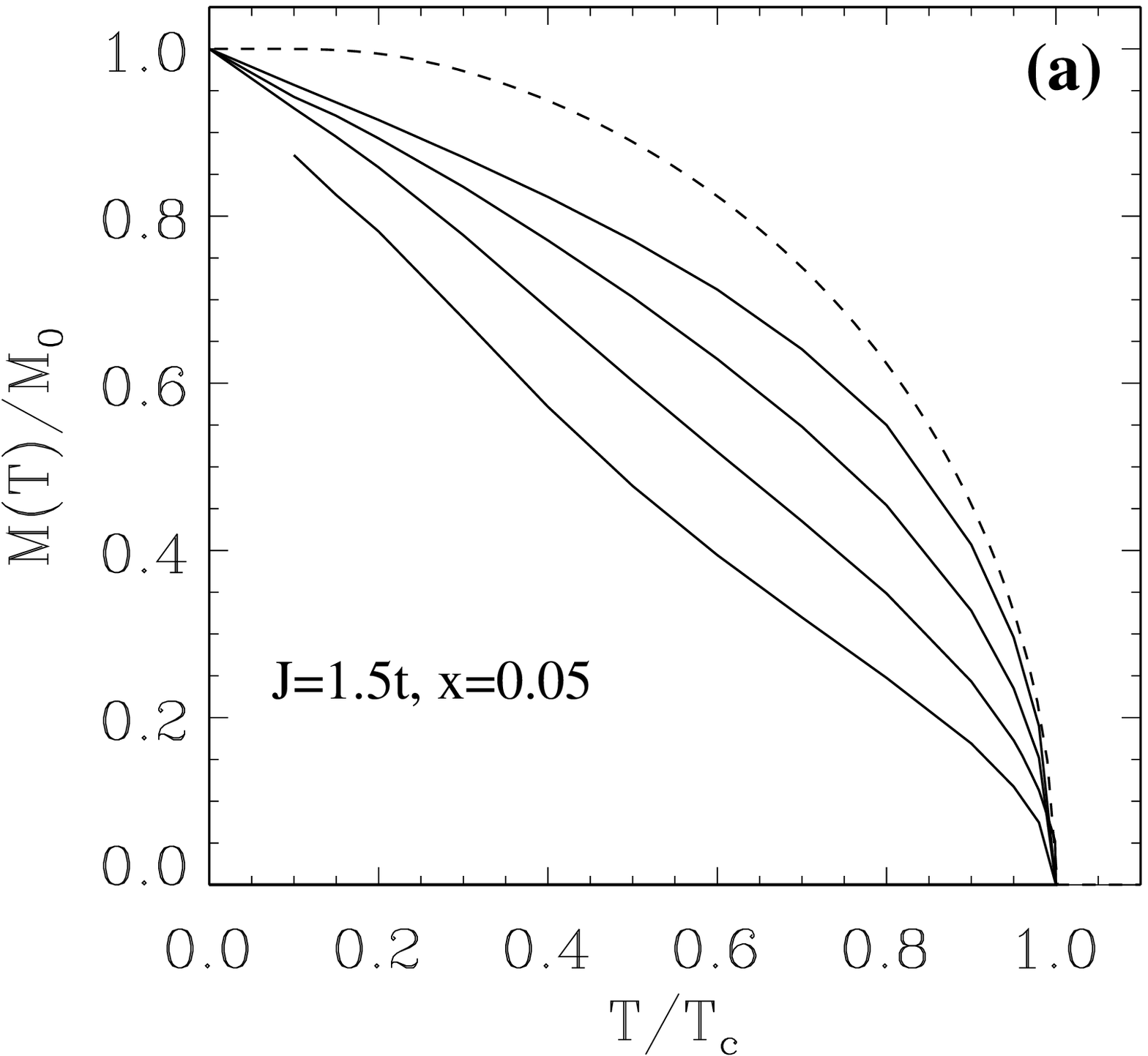}
\includegraphics[width=2.5in]{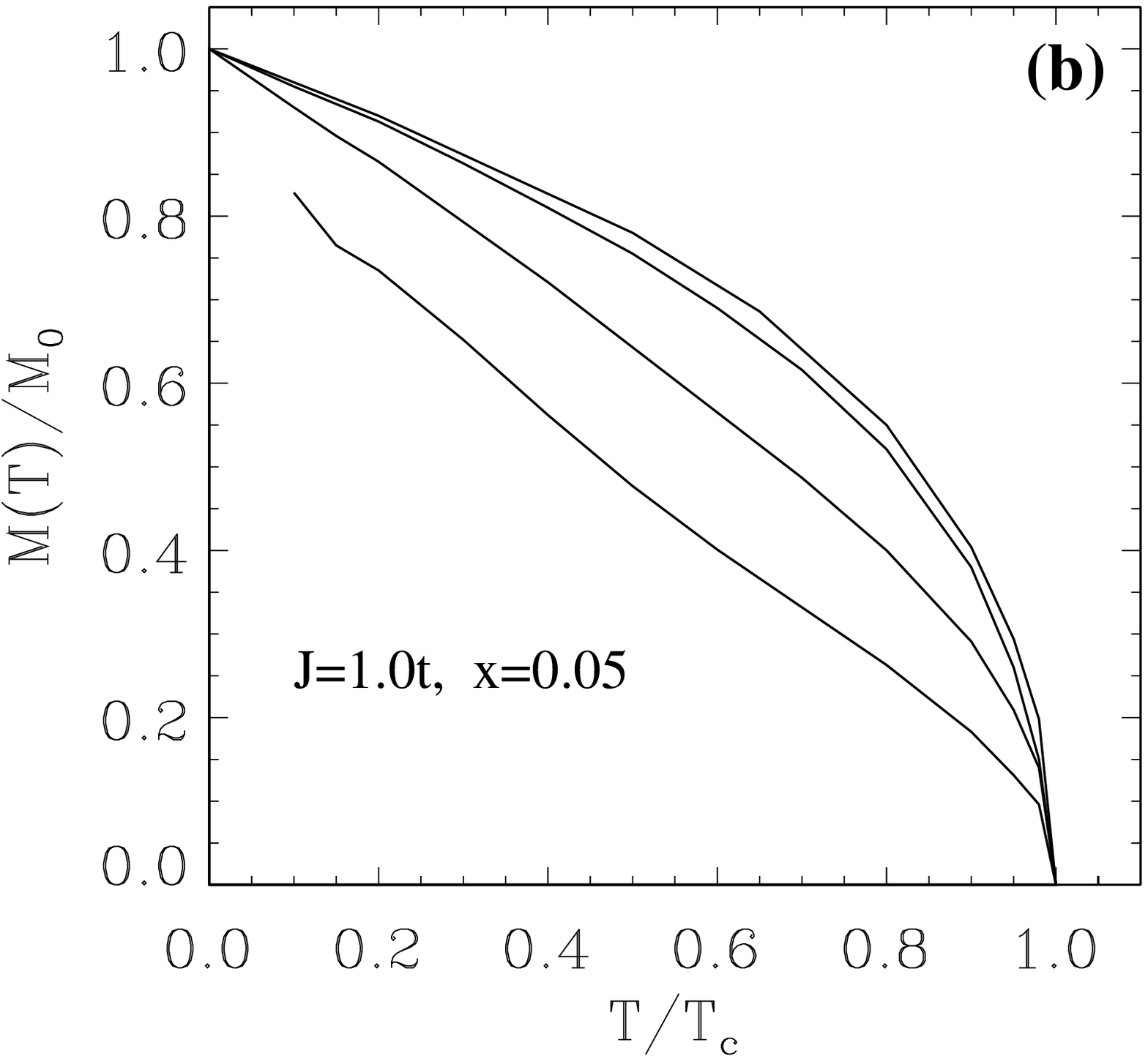}
\includegraphics[width=2.5in]{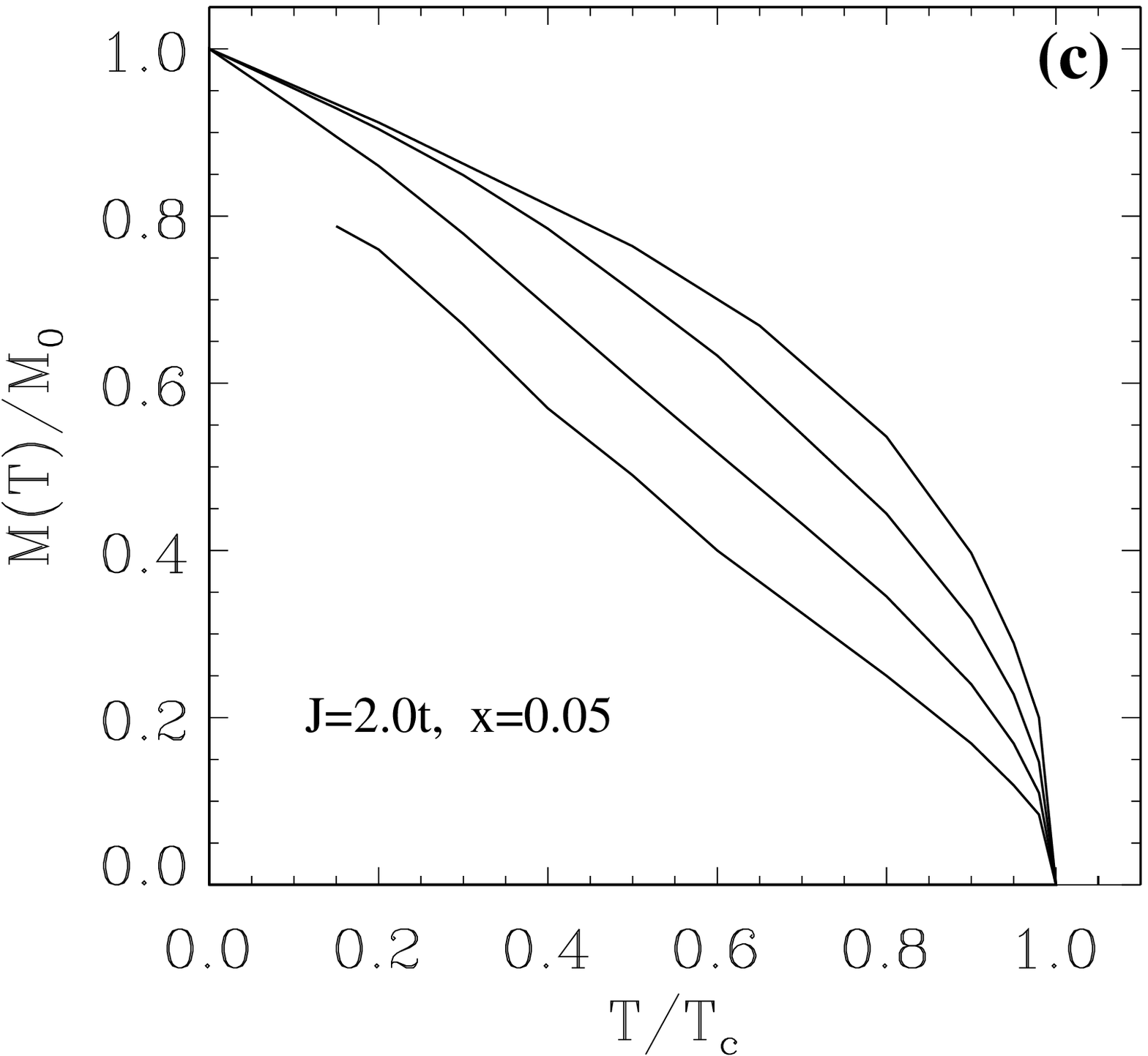}
\caption{
\label{DMFT_fig8}
The normalized DMFT impurity magnetization 
as a function of temperature for (a) $J=1.5t$, (b) $J=1.0t$, and (c)
$J=2.0t$, for $x=0.05$ and for  
$n_{\textrm{c}}/n_{\textrm{i}}=$0.4, 0.2, 0.1, 0.04 (from
the top). 
The dashed line in (a) represents the magnetization calculated 
for the simple Weiss mean field theory for the local moment spin $S=5/2$.
}
\end{figure}

Before we show our calculated magnetization we describe the density of
state (DOS) of dynamical mean-field calculations applying to simple
semi-circle models.  In Fig.~\ref{DMFT_fig7} we show the DMFT density of
states for majority spin near  the band edge corresponding to the
disordered spin state $T>T_c$, and the inset shows the $T=0$
ferromagnetic state.  The evolution of the energy ($\omega$) dependent
DOS is shown as the carrier-spin coupling $J$ is increased from zero;
note that the method works equally well in the ordered $T=0$ state and
the disordered spin $T>T_c$ case, and predicts the formation of a spin
polarized impurity band for $J>t$.  For small $J$ we see the expected
band shift proportional to $xJ$. For $J>J_{c}$ an impurity band centered
at $\sim -J$ and containing $x$ states is seen to split off from the
main band, where the critical coupling $J_c \sim t$. In the DMFT
calculations we parametrized the exchange coupling $J$ in terms of the
band width parameter $t$, subsuming the unit cell volume $a^3_0$
implicitly. 

In Fig.~\ref{DMFT_fig8} we show the normalized magnetization of the
local moments as a function of temperature for different values of
$J=$1.0, 1.5, 2.0$t$ and $x=0.05$, and
for various hole densities, $n_{\textrm{c}}/n_{\textrm{i}}$. 
For the small coupling constant $J=t$
the impurity band is not formed, but for $J=$1.5, 2.0$t$ we have a
spin polarized impurity band.
For relatively high density ($n_{\textrm{c}}/n_{\textrm{i}}=0.4$, or
$n=0.02$) the 
calculated magnetization looks similar to the Weiss mean-field results.
But for low
density ($n_{\textrm{c}}/n_{\textrm{i}}=0.04$) we have a linear
$M(T)$ in the 
intermediate temperature range.  Near the critical temperature $T_c$
the critical behavior of the magnetization for all density is
given by $M(T) \propto (T_c-T)^{1/2}$. For different exchange couplings
we have similar results (\emph{i.\ e.}, linear behavior
at low densities and intermediate temperature ranges).

\begin{figure}
\includegraphics[width=3in]{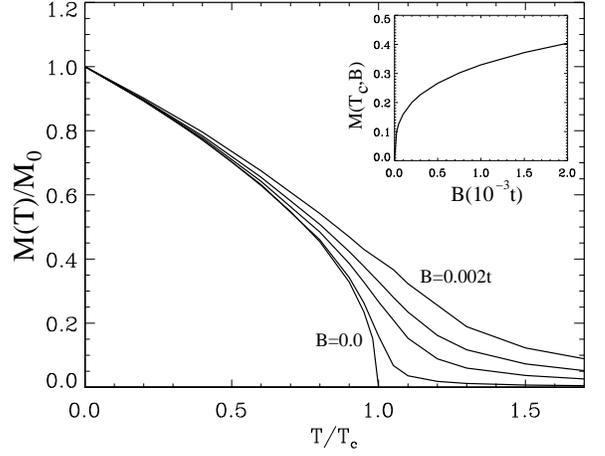}
\caption{
\label{DMFT_fig9}
The external magnetic field effects on the  magnetization 
as a function of temperature for fixed 
parameter values $J=1.5 t$, $x=0.05$, and $n_{\textrm{c}}/n_{\textrm{i}}=0.2$. 
The curves correspond to B=0.002, 0.001, 0.0005, 0.0001, 0.0t 
(from the top).
Inset shows the magnetization as a function of external field at $T=T_c$.}
\end{figure}

In discussing our DMFT magnetization calculations we note that the
DMFT results shown in Fig. \ref{DMFT_fig8} are qualitatively roughly
similar to the delocalized static (degenerate) mean-field-theory
results we obtained in Sec. \ref{sec:2a2}. For example, the results of
Fig. \ref{DMFT_fig8} approximately resemble those shown in
Fig. \ref{mft_fig4}(a) except that the DMFT results of
Fig. \ref{DMFT_fig8} are in much better agreement with the
magnetization measurements in metallic GaMnAs systems in the sense
that the linear behavior of $M(T)$ for lower $T$ (with almost a kink
just below $T_c$ as can be seen in Fig. \ref{DMFT_fig8}) is much more
pronounced (as in the experimental data) than our delocalized
mean-field-theory results shown in Fig. \ref{mft_fig4}(a). This is
both gratifying and expected because DMFT is a substantial improvement
on the Weiss MFT as it incorporates the physics of spin-split impurity
bands through the appropriate quantum self-energy corrections not
included in the simple MFT of Sec. \ref{sec:2a}. In fact, for very low
carrier densities we obtain outwardly concave $M(T)$ in our DMFT
calculations (as we expect to do in the strongly nondegenerate limit),
but the computational convergence in our DMFT numerical calculations
is rather poor in this regime of unrealistically low ($n_c/n_i \le
0.01$) carrier densities, and therefore we refrain from showing these
results. One can, however, detect very slight concavity in the lowest
carrier density ($n_c/n_i = 0.04$) $M(T)$ results shown in
Fig. \ref{DMFT_fig8}. We point out that the critical magnetic
properties in DMFT are the same as in the static MFT, and therefore
all the DMFT critical exponents are equal to those in the Weiss MFT.

In particular, it should be possible to obtain the $M(T)$ behavior for
the localized carrier case ({\it cf.} Sec. \ref{sec:2a1}) also from
DMFT by incorporating impurity band localization in the DMFT
formalism. Our current theory does not include localization, and the
impurity band (or valence band) carriers in our DMFT calculations are
all delocalized metallic carriers. First principles band theory
calculations indicate that the actual exchange coupling in GaMnAs may
be close to critical $J_c$, and as such impurity band physics may be
quite important for understanding DMS magnetization. To incorporate
physics of localization one needs to include disorder effects
(invariably present in real DMS systems) in the model. All our MFT
calculations (both DMFT in Sec. \ref{sec:2b} and the static MFT of
Sec. \ref{sec:2a}) are done in the virtual crystal approximation where
effective field is averaged appropriately leaving out random disorder
effects explicitly. In the next section we explicitly incorporate
disorder in the theory by developing a percolation theory approach to
DMS magnetization for the strongly localized insulating systems.

The calculated magnetization as a function of temperature is shown in
Fig.~\ref{DMFT_fig9} for various external magnetic field values.
The inset shows the magnetization as
a function of external 
field at $T=T_c$.  Our calculated $M(T,B)$ behavior is roughly
qualitatively similar to the mean-field results in section
\ref{sec:2a2}.  At the critical temperature $T_c$ $M(T_c,B)$ shows a
mean-field behavior, $M(T_c,B) \propto B^{1/3}$.
The DMFT results shown in Fig. \ref{DMFT_fig9} are qualitatively
similar to DMS experimental results.

\subsection{Magnetization in the percolation formalism}\label{sec:2c}

Our recently developed percolation theory\cite{sds7} applies strictly
in the regime of strongly localized holes where the dynamical
mean-field theory for delocalized carriers described in
Sec.~\ref{sec:2b} has little validity.  These two theories, mean-field
theory and percolation theory, are therefore complementary.
Interestingly, however, the perturbation theory and the dynamical
mean-field theory are not mutually exclusive in spite of their regimes
of validity being different, and in particular, a significant aspect
of our percolation approach is its ability to reproduce qualitatively
the mean-field theory results of the last section both for $T_c$ and
$M(T)$.

The percolation theory assumes the same carrier-mediated
ferromagnetism model of Sec.~\ref{sec:2a}, but now the carriers are
pinned down with the localization radius $a_B$. The localized carriers
are therefore taken to be in the impurity band and disorder,
completely neglected in the mean-field theory of Sec.~\ref{sec:2a} and
\ref{sec:2b}, now plays a key role in the carrier localization. The
mean-field theory and the perturbation theory are therefore
complementary in the sense that one (the mean-field theory) completely
neglects disorder, and the other (the percolation theory) includes
disorder at a very fundamental level \emph{i.~e.} by starting from the
picture of localized carriers in a strongly disordered system. The
magnetic impurities in this case are assumed to be completely randomly
distributed in the host semiconductor lattice in contrast to the
mean-field case where the carrier states are free and the disorder is
neglected.

As we have demonstrated in Ref.~\onlinecite{sds7}, the
problem of ferromagnetic transition in a system of bound magnetic
polarons can be rigorously reduced to the problem of overlapping
spheres studied in the percolation theory.\cite{EfrosShklovskiiBook}
The latter problem studies spheres of the same radius $r$ randomly
placed in space (three-dimensional in our case) with some
concentration $n$. Overlapping spheres form ``clusters;'' as the
sphere radius $r$ becomes larger, more and more spheres join into
clusters, the clusters coalesce, and finally at some critical value of
the sphere radius an infinite cluster spanning the whole sample
appears.  This problem has only one dimensionless parameter,
$r^3n$, and therefore can be easily studied by means of
Monte-Carlo simulations.

Each sphere of the overlapping spheres problem corresponds to a bound
magnetic polaron, which is a complex formed by one localized hole and
many magnetic impurities with their spins polarized by the exchange
interaction with the hole spin. The concentration $n$ of spheres is
therefore equal to the concentration $n_{\textrm{c}}$ of localized holes. The
expression for the effective polaron radius is not trivial and has
been found in our earlier work.\cite{sds7} The resulting
formal relation between the physical parameters of the system under
consideration and the only parameter of the overlapping spheres
problem reads:\cite{sds7}
\begin{equation}
\label{params}
r^3n = \left[0.86+\left(a_B^3 n_{\textrm{c}}\right)^{\frac{1}{3}}
\ln \frac{T_c}{T}\right]^3\;.
\end{equation}

\begin{figure}
\includegraphics[width=3in]{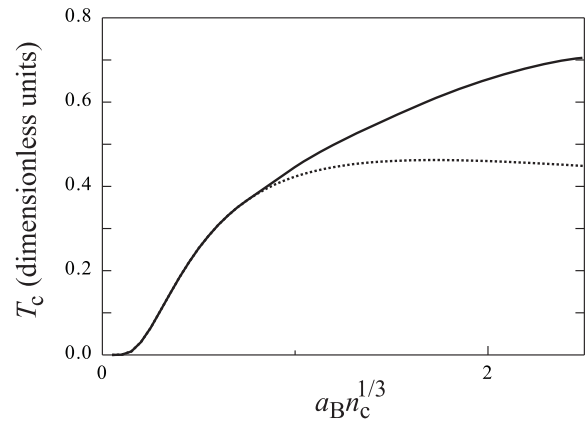}
\caption{\label{fig:Tc_vs_aBnh} Curie temperature $T_c$ as a function of the 
  dimensionless parameter $a_B^3 n_{\textrm{c}}$. At $a_B^3
  n_{\textrm{c}}\lesssim 1$, $T_c$ is given by Eq.~(\protect\ref{Tc}).
  At $a_B^3 n_{\textrm{c}}\gtrsim 1$, Eq.~(\protect\ref{Tc}) (being
  beyond limits of its applicability ) predicts decline of $T_c$
  (dotted line); in reality, $T_c$ grows monotonically with $a_B^3
  n_{\textrm{c}}$ (solid line), though its exact behavior is unknown.
}
\end{figure}

Here $0.64\approx 0.86^3$ is the critical value of the parameter
$r^3n$ at which the infinite cluster appears, and $T_c$ is the
Curie temperature of the ferromagnetic system under consideration,
derived in Ref.~\onlinecite{sds7},
\begin{equation}
\label{Tc}
T_c\sim sSJ\left(\frac{a_0}{a_B}\right)^3  \left(a_B^3n_{\textrm{c}}\right)^\frac{1}{3}\! \sqrt{\frac{n_{\textrm{i}}}{n_{\textrm{c}}}}\
\exp\left(-\frac{0.86}{\left(a_B^3n_{\textrm{c}}\right)^\frac{1}{3}}\right)\;.
\end{equation}
The limit of
applicability of Eq.~(\ref{Tc}) is determined by the condition $a_B^3
n_{\textrm{c}} \ll 1$. The dependence of the Curie
temperature on the hole concentration in a wider domain of values of  
parameter $a_B^3 n_{\textrm{c}}$ is shown schematically in Fig.~\ref{fig:Tc_vs_aBnh}.

\begin{figure}
\includegraphics[width=3in]{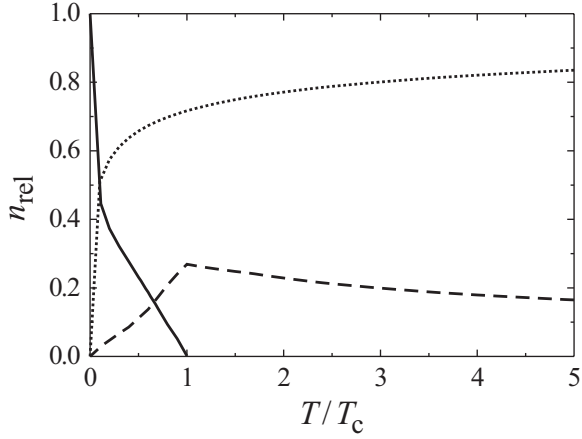}
\caption{\label{fig:v_vs_t} Fractions of volume taken by the infinite
  cluster (solid line) and finite clusters (dashed line) and the
  fraction of volume that do not belong to any cluster or sphere
  (dotted line) for $a_B^3 n_{\textrm{c}} = 10^{-3}$ (Monte-Carlo simulation)}
\end{figure}

Since the magnetic characteristics of the sample are mostly due to
magnetic impurities rather than holes, the quantity of interest is the
number of magnetic impurities in a cluster, which is proportional to
the cluster volume. The magnetic properties of the system can be
expressed in terms of the following quantities, which can be easily
found: (\emph{i}) concentration $\mathcal{P}(vn; r^3n)dv$ of clusters
with volume between $v$ and $v+dv$, (\emph{ii}) the fraction of volume
${P}_\infty (r^3n)$ taken by the infinite cluster, and (\emph{iii})
the fraction ${P}_0 (r^3n)$ which does not belong to any sphere or
cluster of spheres. Clearly, these quantities obey the relation
\begin{displaymath}
{P}_\infty (rn^{\frac{1}{3}}) 
+ \int_0^\infty \mathcal{P}(vn; rn^{\frac{1}{3}})\ v\ dv  
+ {P}_0 (rn^{\frac{1}{3}}) \equiv 1\;.
\end{displaymath}
Figure~\ref{fig:v_vs_t} shows the behavior of the three terms of the
above equation as a function of temperature $T$, with $r^3n$
related to $T/T_c$ by Eq.~(\ref{params}) and $a_B^3 n_{\textrm{c}} = 10^{-3}$.

Having found these quantities, one can easily find the magnetic
properties of the system.

\begin{figure}
\includegraphics[width=3in]{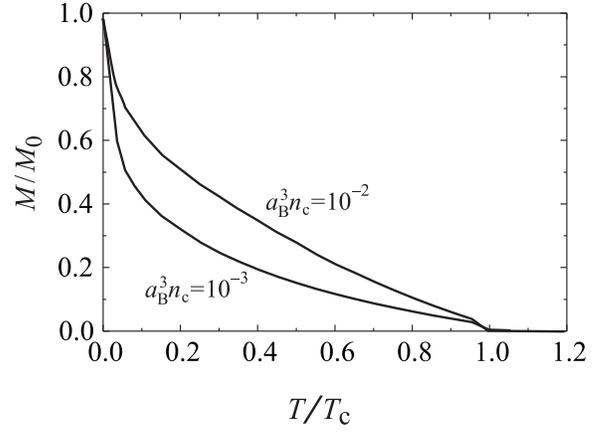}
\caption{\label{fig:m0_vs_t} Spontaneous magnetization as a function
  of temperature.}
\end{figure}

The spins of non-connected
clusters are not correlated and average out when we calculate the
spontaneous magnetization of the whole sample. The only non-vanishing
contribution comes from the infinite cluster. The total magnetic
moment of the sample per unit volume is therefore
\begin{equation}
\label{M0}
M = n_{\textrm{i}} S {P}_\infty \left(\frac{T}{T_c}, a_B^3 n_{\textrm{c}}\right)\;.
\end{equation}
Here and in the following equations we write the characteristics
$P_\infty$, $P_0$, $\mathcal{P}(vn)$ of the overlapping sphere problem
as functions of the physical parameters $T/T_c$ and $a_B^3
n_{\textrm{c}}$ instead of $r^3n$. The relation between these
parameters is given by Eq.~(\ref{params}). The temperature dependence
of the spontaneous magnetization given by Eq.~(\ref{M0}) is plotted on
Fig.~\ref{fig:m0_vs_t} for two experimentally viable values of $a_B^3
n_{\textrm{c}}$. As already mentioned before, such strongly concave
$M(T)$ behavior is often observed in insulating DMS systems, see for
example Refs.~\cite{new4,new6,new38}

\begin{figure}
\includegraphics[width=3in]{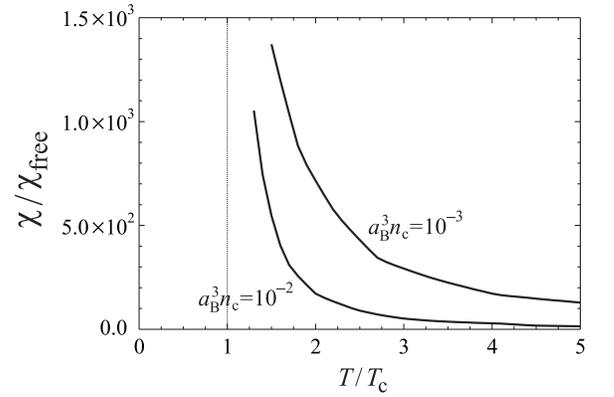}
\caption{\label{fig:chi_vs_t} Temperature dependence of the magnetic
  susceptibility for $n_{\textrm{c}}/n_{\textrm{i}}=0.02$.} 
\end{figure}

The magnetic susceptibility of a sample has contributions coming from
polaron clusters and free spins. Using the classical expression for
the susceptibility of a free magnetic moment $m_0$
\begin{displaymath}
\chi_{m_0}=\frac{m_0^2}{3T}
\end{displaymath}
and the fact that the total spin of a cluster with volume $v$ equals
$vn_{\textrm{i}}$ we arrive at
\begin{eqnarray}
\chi = \frac{1}{3T} gS^2 \left[
\int_0^\infty \mathcal{P}\left(vn_{\textrm{c}}; \frac{T}{T_c}, a_B^3 n_{\textrm{c}}\right)\ 
(vn_{\textrm{i}})^2\ dv
\right.\nonumber\\
\left.\vphantom{\int_0^\infty}
+n_{\textrm{i}} {P}_0 \left(\frac{T}{T_c}, a_B^3 n_{\textrm{c}}\right)
\right]\;.
\label{chi}
\end{eqnarray}
The first term in brackets diverges as $(T-T_c)^\gamma$ with
$\gamma\approx 1.7$ at $T\to T_c$. The second term in brackets does
not have any singularity at $T\to T_c$.  For the temperature
dependence of the susceptibility given by Eq.~(\ref{chi}), see
Fig.~\ref{fig:chi_vs_t}.

\begin{figure}
\includegraphics[width=3in]{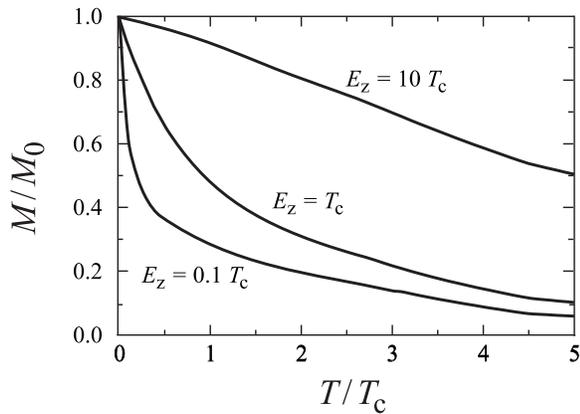}
\caption{\label{fig:m_vs_t} Temperature dependence of magnetization in
  finite magnetic field with $E_Z=g_{\textrm{i}}\mu_BSB$.}
\end{figure}

\begin{figure}
\includegraphics[width=3in]{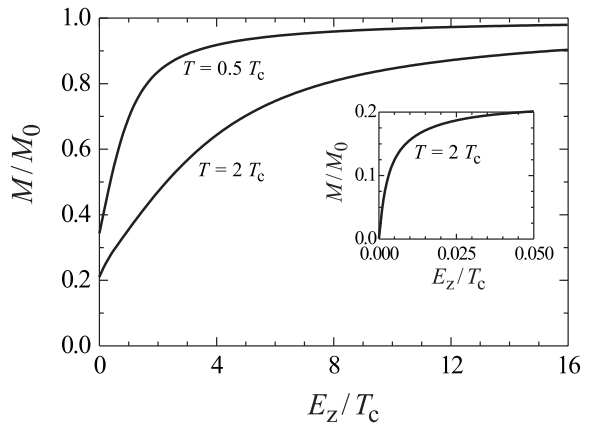}
\caption{\label{fig:m_vs_b} Magnetic field dependence of the sample
  magnetization at given temperature with $E_Z=g_{\textrm{i}}\mu_BSB$.
}
\end{figure}

Using the classical magnetization relation
\begin{equation}
\label{langevin}
M_S(T) = S \mathcal{L}\left(\frac{g_{\textrm{i}}\mu_BSB}{T}\right)\;,
\end{equation}
where $\mathcal{L}(x)\equiv\mathcal{B}_\infty(x)=  \mathrm{cotan}\ x - 1/x$ is the Langevin
function, we obtain the expression for the magnetic moment per unit
volume in finite magnetic field 
\begin{eqnarray}
M(B) &=& n_{\textrm{i}} S {P}_\infty \left(\frac{T}{T_c}, a_B^3 n_{\textrm{c}}\right)\nonumber\\
&+&\!\!\int_0^\infty\!\! \mathcal{P}\left(vn_{\textrm{c}}; \frac{T}{T_c}, a_B^3
n_{\textrm{c}}\right)\ vn_{\textrm{i}}\  
\mathcal{L} \left(\frac{g_{\textrm{i}}\mu_Bvn_{\textrm{i}}SB}{T}\right)dv\nonumber\\
&+&n_{\textrm{i}} S P_0\left(\frac{T}{T_c}, a_B^3 n_{\textrm{c}}\right)
\mathcal{L} \left(\frac{g_{\textrm{i}}\mu_BSB}{T}\right)\;.
\label{MB}
\end{eqnarray}
Figures~\ref{fig:m_vs_t} and \ref{fig:m_vs_b} illustrate the
dependence of the magnetization on the temperature and magnetic field
respectively.  Since the spins of polarons and their clusters are much
large than those of three impurities, the magnetization curve of
Fig.~\ref{fig:m_vs_b} has two characteristic scales. First, polarons
and their clusters are polarized (inset of Fig.~\ref{fig:m_vs_b}),
then the free spins are polarized at much larger fields.

We mention that our percolation-theory critical exponents are
$\gamma\approx 1.7$ and $\beta\approx 0.4$ for the susceptibility
(Fig.~\ref{fig:chi_vs_t}) and magnetization (Fig.~\ref{fig:m0_vs_t}),
as compared with the mean-field results of $\gamma=1$ and $\beta=0.5$
and the best existing numerical estimates of $\gamma\approx1.39$ and
$\beta\approx 0.37$ for the three-dimensional Heisenberg model.\cite{new39}

\section{Conclusion}\label{sec:3}

The temperature-dependent magnetization results presented in this work
for different but interrelated theoretical models are qualitatively
consistent with one another. In particular, the two theories for
metallic DMS systems with itinerant carriers, namely the
degenerate-carrier Weiss static mean-field theory (see
Sec.~\ref{sec:2a2}) and the dynamical mean-field theory (see
Sec.~\ref{sec:2b}), both give onwardly convex $M(T)$ carriers, similar
(but not identical) to that in the textbook molecular mean-field
result, with the convexity being enhanced (suppressed) with increasing
(decreasing) delocalized carrier density. In typical situations
involving itinerant carriers in metallic DMS systems, we find
temperature dependent magnetization curves which are almost linear at
lower temperatures in excellent agreement with experimental
observations in metallic GaMnAs systems.  Such apparent
``non-mean-field-like'' magnetization behavior, found here both in our
delocalized-carrier static and dynamic mean-field theories, arises
from a combination of reasons: (1) the double Brillouin function
[\emph{i.~e.} $\mathcal{B}(\mathcal{B}(x))$] form of the coupled field
magnetization [see Eq.~(\ref{sieq})]; (2) the low values of
$n_{\textrm{c}}/n_{\textrm{i}}$, leading to the Mn moments feeling on
the average a much reduced number of free carriers. As the carrier
density is increased, for example by suitable annealing in recent
experimental studies, such linear magnetization curves evolve toward
the more conventional outwardly convex magnetization, as can be seen
in our results and in recent annealing
experiments.\cite{new15,new16,new17,new18,new19} In the case of
localized carriers appropriate for insulating DMS systems (many of
which are also found to be ferromagnetic with well-defined Curie
temperatures, \emph{e.~g.}  Ga$_{1-x}$Mn$_x$As for $x<0.03$,
In$_{1-x}$Mn$_x$As,\cite{new4} and Ge$_{1-x}$Mn$_x$\cite{new6}), our
two complementary theories, namely the non-degenerate-carrier static
mean-field theory (Sec.~\ref{sec:2a1}) and the percolation theory
(Sec.~\ref{sec:2c}), give qualitatively similar magnetization
behavior. In particular, at low carrier densities the $M(T)$ curves in
strongly insulating DMS systems exhibit strikingly
``non-mean-field-like'' outwardly concave magnetization behavior, as
often observed in insulating DMS systems.  Again, this unusual concave
magnetization behavior arises from a combination of the strongly
localized nature of the carrier system and the low values of the
carrier density (\emph{i.~e.}  $n_{\textrm{c}}/n_{\textrm{i}}\ll 1$)
in DMS materials. We emphasize, however, a very important feature of
theoretical results presented in this work: the convex magnetization
behavior in the strongly metallic case and the concave behavior in the
insulating case are \emph{not} a sharp dichotomy in DMS properties ---
these are really the two extremes of a continuum of possible
magnetization behavior in DMS materials, as can be inferred from the
results of Sec.~\ref{sec:2a}. The typical DMS magnetization behavior
should lie somewhere in between these two extremes of highly concave
(low carrier density and strongly localized insulating DMS) and highly
convex (high carrier density and strongly delocalized metallic DMS)
magnetization, leading to the generic experimental observation of
almost linear $M(T)$ behavior in ferromagnetic Ga$_{1-x}$Mn$_x$As.

Another often-mentioned peculiar aspect of DMS magnetization, namely
the low value of the saturation magnetization (compared with the
number of Mn ions in the sample as given by the value of $x$ in
Ga$_{1-x}$Mn$_x$As), is also apparent in our theoretical results
presented in this work. In particular, the presence of direct
antiferromagnetic coupling between the Mn moments may drastically
suppress the saturation magnetization, as can be seen in the results
presented in Figs.~\ref{mft_fig2} and \ref{mft_fig5} of this paper.
This absence of complete magnetization saturation in our mean-filed
results is akin to having effective ferrimagnetism in the system.  In
addition, the calculated magnetization is may be much lower than the
saturation value except at very low temperatures, see
Fig.~\ref{fig:m0_vs_t}, which could also explain the lack of
magnetization saturation. It is likely that in the real DMS materials
an appreciable fraction of the Mn moments are magnetically inactive
because they do not sit in cation substitutional sites, but rather at
defect sites such as interstitials and are antiferromagnetically
coupled to other Mn moments. Such magnetically inactive Mn atoms could
be one of the reasons for the low values of saturation magnetization
in Ga$_{1-x}$Mn$_x$As. We also note that our percolation theory
provides another possible explanation for the observed low values of
saturation magnetization in low carrier density insulating DMS
materials. Since the infinite cluster of percolating bound magnetic
polarons triggering the long-range ferromagnetic ordering necessarily
leaves out a large number of Mn moments (which are not parts of the
infinite cluster except at $T\to 0$), one naturally expects a very low
saturation magnetization except perhaps at $T\ll T_c$.

A recent series of potentially important annealing experiments in
metallic Ga$_{1-x}$Mn$_x$As samples by several different groups may
eventually shed considerable light on our understanding of the
mechanisms underlying DMS ferromagnetism. These experiments, while
differing somewhat on the details, all find that suitable low
temperature ``optimal'' annealing may enhance the magnetic properties
of Ga$_{1-x}$Mn$_x$As by increasing $T_c$ and more importantly for our
purpose, by enhancing $M(T)$ to more convex almost standard
mean-field-like behavior. This enhancement of magnetization seems to
correlate well with improvement in the metallicity of the annealed
sample (\emph{e.~g.} higher conductivity) and with an increase in the
hole density. Annealing may also be enhancing the magnetic properties
by annealing away some of the Mn interstitials. Our theoretical
results are in excellent agreement with these annealing experiments,
since increasing $n_{\textrm{c}}/n_{\textrm{i}}$ does lead to enhanced
(and more convex) magnetization in our theory. In this context, it
will be very helpful to have more detailed information on the $M(T)$
behavior in strongly localized insulating samples both for
Ga$_{1-x}$Mn$_x$As with $x<0.03$ and for other insulating DMS
materials (\emph{e.~g.}  Ge$_{1-x}$Mn$_x$).

Our final comment addresses the role of disorder (essentially
neglected in our work, except for the percolation theory part, as we
treat it in the virtual crystal approximation, \emph{i.~e.}, we assume
that holes and local moments only feel the average effective field ),
which is invariably strong (and not yet well understood) in DMS
materials. In particular, even the ``metallic'' DMS systems are in
effect very poor metals with mean-free paths which are at or below the
Ioffe-Regel limit (with low-temperature mobilities of the order of a
few cm$^2$/V$\cdot$s only), indicating the presence of very strong
disorder.\cite{new41} In the presence of such strong disorder, various
spin glass ground states may compete with ferromagnetic ground states.
We believe that it is important to look for signatures of spin glass
physics in low-temperature DMS magnetic properties. It may very well
turn out that spin glass phases dominate the regime of parameter space
(\emph{e.~g.} $x<1$\% or $x>10$\% in Ga$_{1-x}$Mn$_x$As) where a
ferromagnetic ground state does not seem to stabilize in DMS
materials.

This work is supported by US-ONR and DARPA.

\end{document}